\documentclass[pdflatex,sn-nature]{sn-jnl}

\usepackage{graphicx}
\usepackage{multirow}
\usepackage{amsmath,amssymb,amsfonts}
\usepackage{amsthm}
\usepackage{mathrsfs}
\usepackage[title]{appendix}
\usepackage{xcolor}
\usepackage{textcomp}
\usepackage{manyfoot}
\usepackage{booktabs}
\usepackage{bm}
\usepackage{upgreek}
\usepackage{placeins}
\usepackage{pdfpages}

\theoremstyle{thmstyleone}

\theoremstyle{thmstyletwo}

\theoremstyle{thmstylethree}

\begin{document}

\title[Self-verifying gas analysis from raw sensor signals]{Self-verifying time-resolved multicomponent gas analysis from raw sensor signals}

\author*[1,2]{\fnm{YingCheng} \sur{Zhou}}\email{ZHOU.Yingcheng@nims.go.jp}

\author[1]{\fnm{Kosuke} \sur{Minami}}\email{MINAMI.Kosuke@nims.go.jp}

\author*[1,2]{\fnm{Genki} \sur{Yoshikawa}}\email{YOSHIKAWA.Genki@nims.go.jp}

\affil*[1]{\orgdiv{Research Center for Macromolecules and Biomaterials}, \orgname{National Institute for Materials Science (NIMS)}, \orgaddress{\street{1-1 Namiki}, \city{Tsukuba}, \postcode{305-0044}, \state{Ibaraki}, \country{Japan}}}

\affil[2]{\orgdiv{Materials Science and Engineering, Graduate School of Pure and Applied Science}, \orgname{University of Tsukuba}, \orgaddress{\street{1-1-1 Tennodai}, \city{Tsukuba}, \postcode{305-8571}, \state{Ibaraki}, \country{Japan}}}

\abstract{Gas sensor arrays can estimate the composition of gas mixtures from their response patterns. However, when a sensor is used outside the laboratory, the true gas concentrations are unknown, so there is usually no direct way to determine whether the estimated result is reliable. Here, we develop a physics-constrained neural network that estimates time-varying gas concentrations and reconstructs the corresponding sensor signals at the same time. We tested the method with a twelve-channel membrane-type surface stress sensor array exposed to water--ethanol mixtures whose concentrations changed over time. The model reconstructed the concentration profiles of both components from the raw signals. Artificial offsets and sensitivity changes in one channel were detected from disagreement between the measured and reconstructed signals. Deletion or replacement of part of a measurement record was detected from violations of the sorption and mechanical equations. This approach enables time-resolved multicomponent gas analysis with self-verification from a single measurement.}


\maketitle

Chemical sensing is moving out of the laboratory into everyday use, from environmental monitoring~\cite{zong2024} and industrial process monitoring~\cite{maji2025} to food-quality control~\cite{mor2025} and breath-based health screening~\cite{loizeau2013, saeki2024b}, where a sensor must often work continuously and with little opportunity to recalibrate against a known reference. Equilibrium-type chemical sensors, including quartz crystal microbalances, surface acoustic wave devices, and membrane-type surface stress sensors (MSS), are widely deployed for gas detection owing to their sensitive transduction of gas--solid interactions across diverse receptor materials~\cite{devkota2018, matatagui2019, oon-pitipongsa2026, malhotra2023, minami2025, minami2024}. In many practical implementations, these sensors are operated under carefully programmed input sequences~\cite{imamura2018c, minami2021a}, and their outputs are reduced to static or quasi-static features~\cite{shiba2017a, liu2023f, yan2015}. Such strategies simplify calibration and improve reproducibility. At the same time, dynamic sensor responses have long been used to extract kinetic parameters, identify gases, and estimate changing concentrations~\cite{imamura2018c, minami2021a, imamura2019a, fonollosa2015}. The issue addressed here is therefore not whether transient signals are informative, but whether the full waveform can also provide an internal check on a time-resolved concentration estimate when the true concentration is unknown.

The problem becomes acute when gas compositions vary continuously in time or when multiple components are present simultaneously. Under such conditions, the sensor response at any instant depends on the exposure history, so no single snapshot of the signal determines the composition. Existing approaches have often addressed this difficulty by emphasizing selectivity, for example through material design, array diversity, or stimulus programming that separates responses in time. Many of these strategies remain effective in controlled settings, but they commonly treat gases other than the target as interference to be rejected. The alternative pursued here is to estimate every component that contributes to the signal, so that interference is explained rather than rejected. This route raises a broader challenge: how to recover time-varying gas concentrations from coupled transient responses, and how to judge whether the recovered answer is consistent with what the sensor actually measured.

Recent studies have shown that dynamic sensor responses can support informative gas analysis beyond conventional steady-state operation. Data-driven approaches, including free-hand measurements and reservoir-computing-based inference, have estimated time-varying gas identities and concentrations from transient multichannel responses under less controlled conditions~\cite{imamura2019a, fonollosa2015}. In parallel, analytical inversion based on sorption kinetics has enabled multi-component concentration recovery under programmed stimulation, but typically requires explicit forward models and pre-characterized component-specific parameters~\cite{minami2025, feng2025a}. These studies demonstrate that dynamic responses have already been exploited for gas analysis. What remains difficult in deployment is to evaluate a time-resolved prediction when its ground-truth trajectory is unavailable.

This motivates treating the sensor explicitly as a dynamical system rather than only as a static transducer. When the gas composition changes, molecules sorb into and out of the receptor film at finite rates, the film swells, and mechanical stress builds and relaxes. Each of these processes has its own time scale, and the measured signal traces their combined evolution. The transient waveform is therefore a time-resolved record of the gas exposure that produced it, written in the sensor's own dynamics.

This picture places chemical sensing in a familiar mathematical setting. The sensor is a dynamical system whose internal state, the amount of gas sorbed in each receptor film, cannot be measured directly but leaves its signature in the output. Recovering hidden states and inputs from measured outputs is a classical inverse problem, studied as state estimation in control and data assimilation. We solve it here with a neural network required to obey the governing equations of sorption and viscoelastic relaxation~\cite{karniadakis2021}, a strategy we refer to as constrained dynamical inversion. The constraints do more than regularize. Because the inferred concentrations must reproduce every channel of the measured waveform through the same equations, the inference is redundant by construction, and this redundancy is exactly what can be checked at deployment. Verification stops being an external add-on and becomes a property of the solution.

\begin{figure*}[htbp]
    \centering
    \includegraphics[width=\textwidth]{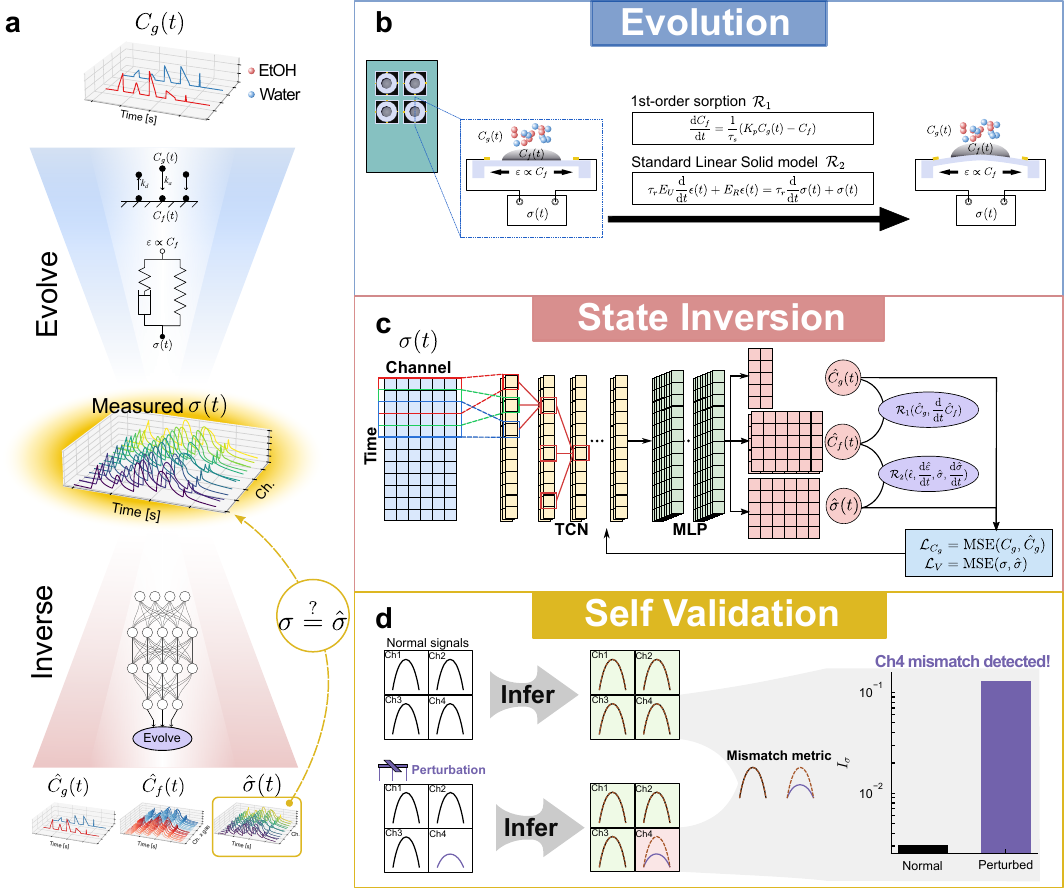}
    \caption{Conceptual overview of the constrained dynamical inversion framework for chemical sensing.
    (a)~End-to-end scheme. A time-varying gas-phase concentration $C_g(t)$ of ethanol (EtOH) and water evolves through the sensor physics into the measured multichannel signals $\sigma(t)$ (Evolve): gas molecules sorb into the receptor films, the film-phase concentration $C_f(t)$ builds up, and the strain $\varepsilon \propto C_f$ generates surface stress. A network inverts the measured signals back into gas-phase concentrations $\hat{C}_g(t)$, hidden film-phase concentrations $\hat{C}_f(t)$, and a reconstruction $\hat{\sigma}(t)$ of the signal itself (Inverse). The same forward model constrains these three outputs (Evolve), so the reconstructed $\hat{\sigma}$ stays consistent with the measured $\sigma$. Their comparison $\sigma \stackrel{?}{=} \hat{\sigma}$ provides self-validation without ground truth.
    (b)~Evolution (forward model). For each receptor on the MSS array, gas at concentration $C_g(t)$ sorbs into the film at concentration $C_f(t)$ following first-order sorption kinetics (residual $\mathcal{R}_1$), and the strain $\varepsilon \propto C_f$ generates the surface stress $\sigma(t)$ through a standard linear solid (SLS) model (residual $\mathcal{R}_2$). Together these define the forward map $C_g \rightarrow C_f \rightarrow \sigma$.
    (c)~State inversion (network). The multichannel signal $\sigma(t)$ (channel $\times$ time) is encoded by a temporal convolutional network (TCN) and decoded by a shared multilayer perceptron (MLP) into three output heads $\hat{C}_g(t)$, $\hat{C}_f(t)$, and $\hat{\sigma}(t)$. The heads are trained against the data losses $\mathcal{L}_{C_g} = \mathrm{MSE}(C_g, \hat{C}_g)$ and $\mathcal{L}_V = \mathrm{MSE}(\sigma, \hat{\sigma})$, subject to the physical residuals $\mathcal{R}_1$ and $\mathcal{R}_2$ that couple the three outputs.
    (d)~Self-validation. Normal signals are inferred into per-channel reconstructions that match the input (green). When a single channel is perturbed (Ch4), its reconstructed $\hat{\sigma}$ can no longer match the corrupted input (red), and the mismatch is quantified by the signal inconsistency $I_{\sigma}$, which rises sharply for the perturbed channel and localizes the fault.}
    \label{fig:conceptual_schematic}
\end{figure*}

Figure~\ref{fig:conceptual_schematic} lays out the approach. In the forward direction, a time-varying gas concentration drives uptake into each receptor film, and the swelling film generates surface stress, producing the measured multichannel signal (Fig.~\ref{fig:conceptual_schematic}a,b). Our framework runs this backward. A single network reads the raw signal and returns three quantities at once (Fig.~\ref{fig:conceptual_schematic}a,c): the gas-phase concentrations $\hat{C}_g$, the hidden film-phase state $\hat{C}_f$ that cannot be measured directly, and a reconstruction $\hat{\sigma}$ of the signal itself. Because the same physics ties these outputs together, $\hat{\sigma}$ must stay consistent with the measured $\sigma$, so their agreement is a check that needs no ground truth (Fig.~\ref{fig:conceptual_schematic}d). When a channel is corrupted, its reconstruction can no longer match the input, and the mismatch both flags the fault and localizes it.

Using a twelve-channel MSS array exposed to dynamic water--ethanol mixtures with several waveform patterns, we first detail the physical constraints that couple the three outputs, then show that the framework recovers time-resolved concentration trajectories from the raw multichannel signals. We then show that the same inference verifies itself. The reconstructed signal detects global corruptions of the input, and the physical residuals detect subtler anomalies that leave the signal statistics intact. Together, these results recast dynamic chemical sensing as constrained state inference rather than static pattern recognition.

\section{Constrained dynamical inversion}\label{sec:physics}

The reconstruction model (Fig.~\ref{fig:conceptual_schematic}c) takes the multichannel time-series signals $\sigma(t)$ as input. A temporal convolutional network (TCN) encodes temporal features, and a shared multilayer perceptron (MLP) decodes them into three output heads at each time step: inferred gas-phase concentrations $\hat{C}_g$, film-phase concentrations $\hat{C}_f$ representing the internal state of the receptor films, and a reconstruction $\hat{\sigma}$ of the measured signal itself. Architectural details are given in Methods (Sections~\ref{subsec:architecture}--\ref{subsec:training}). The three outputs are not independent but coupled through the governing equations of the sensing process, described below, which are enforced as constraints during training so that they remain mutually consistent.

Sorption kinetics governs the uptake of gas molecules into the receptor film~\cite{imamura2018c, minami2021a}:
\begin{equation}\label{eq:sorption}
    \frac{d C_f}{dt} = \frac{1}{\tau_s} (K_p C_g - C_f),
\end{equation}
where $\tau_s$ is the sorption time constant and $K_p$ the partition coefficient. Uptake swells the film, and under the small-strain assumption the resulting strain is proportional to the film-phase concentration, $\hat{\varepsilon} = (v/3)\,\hat{C}_f$, with $v$ the specific volume of the sorbed species. This strain generates the surface stress, whose evolution is governed by a standard linear solid (SLS) model~\cite{heinrich2009,wenzel2008b}:
\begin{equation}\label{eq:sls}
    \tau_r E_U \frac{d \varepsilon}{dt} + E_R \varepsilon = \tau_r \frac{d \sigma}{dt} + \sigma,
\end{equation}
where $E_U$ and $E_R$ are the unrelaxed and relaxed moduli, and $\tau_r$ the mechanical relaxation time constant. These two equations establish a sequential dependence $C_g \rightarrow C_f \rightarrow \sigma$. The gas-phase concentration drives film uptake, which in turn drives the mechanical response. The multi-gas, multi-channel formulation is detailed in Supplementary Note~3.

During training, the governing equations are not solved explicitly but enforced as soft constraints through their residuals:
\begin{align}
    \mathcal{R}_1 &= \frac{d \hat{C}_f}{dt} - \frac{1}{\tau_s}\bigl(K_p \hat{C}_g - \hat{C}_f\bigr), \label{eq:r1}\\
    \mathcal{R}_2 &= \tau_r E_U \frac{d \hat{\varepsilon}}{dt} + E_R \hat{\varepsilon} - \tau_r \frac{d \hat{\sigma}}{dt} - \hat{\sigma}. \label{eq:r2}
\end{align}
The common physics-informed practice adds such residuals as fixed-weight penalty terms in the loss. Because the data and physics objectives compete, this weighted sum can only reach solutions on the convex part of their Pareto front, so no fixed weight recovers a balanced optimum that lies on a non-convex region, and the result depends on manual weight tuning~\cite{almanstotter2025}. We instead impose them with the modified differential multiplier method (MDMM)~\cite{platt,almanstotter2025}, which treats the residuals as constraints and converges to the saddle point of the Lagrangian that constitutes the constrained optimum, adapting the multipliers automatically during training. This distinction matters because the physics must actually hold for the reconstruction to serve as an independent consistency check, not merely regularize it. Under MDMM, $\mathcal{R}_1$ and $\mathcal{R}_2$ act as constraints rather than loss terms. The optimizer minimizes the data losses $\mathcal{L}_{C_g} = \mathrm{MSE}(C_g, \hat{C}_g)$ and $\mathcal{L}_V = \mathrm{MSE}(\sigma, \hat{\sigma})$ subject to these residuals remaining small. This forces the network to produce trajectories of $\hat{C}_f$, $\hat{C}_g$, and $\hat{\sigma}$ that are mutually consistent under the assumed dynamics. The film-phase concentration $\hat{C}_f$ receives no direct experimental supervision and is determined entirely by these constraints.

\section{Concentration reconstruction and joint physical consistency}\label{sec:reconstruction}

We assess whether this framework can recover time-varying gas concentrations from raw multichannel sensor signals acquired across multiple input waveforms. Experimental data were acquired with a home-built membrane-type surface stress sensor array comprising three four-channel chips (twelve channels total) coated with twelve chemically distinct receptor layers to span a diverse set of sorption and viscoelastic responses. The framework itself does not depend on this specific channel count. Dynamic binary vapor mixtures of water and ethanol were generated by bubbling N$_2$ through the respective liquids and mixing the resulting vapors. This pairing is not incidental. Water vapor is the dominant interferent in most real gas-sensing deployments, and here it is inferred on an equal footing with ethanol, so its contribution to the signal is accounted for rather than compensated away. Concentration amplitudes, waveform types, and segment durations were varied among rectangular, sawtooth, reverse sawtooth, and Gaussian patterns. The dataset comprises two regimes: synchronous mixing, in which both components share a common waveform, and asynchronous mixing, in which each component follows an independently generated waveform. In total, 150 sequences were used for training and 50 held out for testing (Methods, Section~\ref{subsec:gasdelivery}). All network parameters were frozen after training convergence.

\begin{figure*}[htbp]
    \centering
    \includegraphics[width=\textwidth]{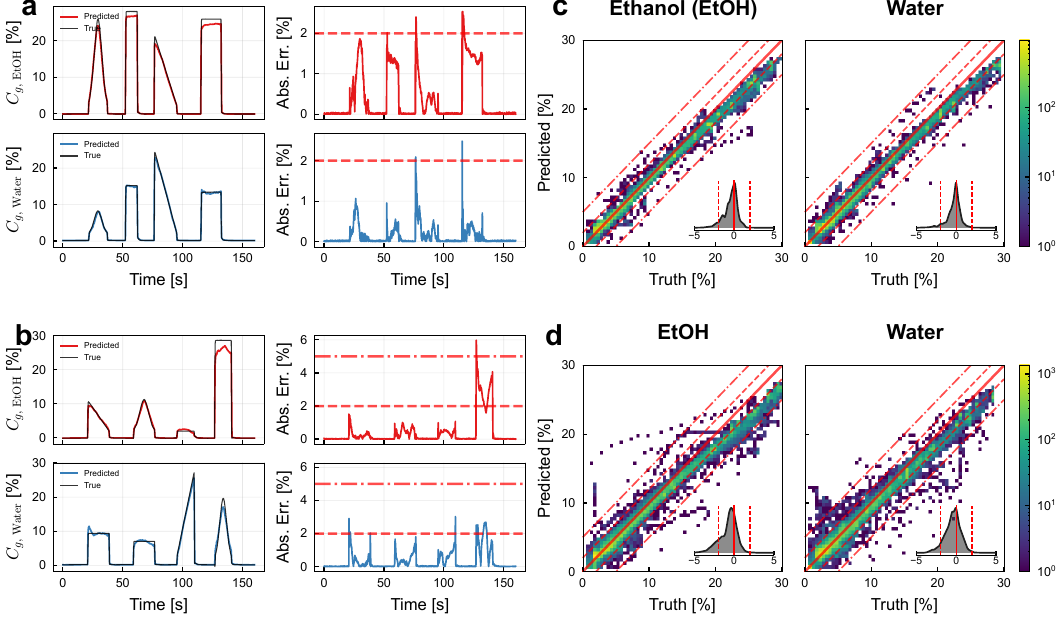}
    \caption{Reconstruction of mixed-gas concentration trajectories from variable dynamic responses.
    (a)~Synchronous mixing: reconstructed (colored) and ground-truth (black solid) concentration trajectories for ethanol (top left) and water (bottom left), with the corresponding absolute errors (right panels).
    (b)~Asynchronous mixing, same layout as (a).
    (c,~d)~Two-dimensional density plots of reconstructed versus ground-truth concentrations for ethanol (left) and water (right) under synchronous~(c) and asynchronous~(d) conditions. Color encodes frequency. The red diagonal indicates perfect agreement, with dashed lines marking $\pm$2\% and dash-dotted lines marking $\pm$5\% deviation. Points below 0.5\% in both predicted and true concentrations are excluded. Insets show the kernel density estimate (KDE) of the overall reconstruction error distribution.}
    \label{fig:reconstruction}
\end{figure*}

Figure~\ref{fig:reconstruction}a,b shows that the framework recovers the full temporal structure of the concentration trajectories under both synchronous and asynchronous mixing: peak positions, waveform shapes, and relative amplitudes between species are faithfully reproduced from the raw multichannel signals alone. The largest deviations occur near sharp rising and falling edges, where rapid concentration changes outpace the sensor's sorption response and compress distinguishing features in the transient signal. This is notable because rectangular (step-function) stimuli, which maximize such sharp transitions, are the most widely used gas-delivery protocol in laboratory settings. The present results suggest that these abrupt waveforms, while convenient for calibration, are not necessarily optimal for dynamic trajectory reconstruction.

Correlation plots and marginal error distributions (Fig.~\ref{fig:reconstruction}c,d) show that per-component errors are concentrated within approximately $\pm$2 percentage points under synchronous and $\pm$5 percentage points under asynchronous conditions, with a slight systematic deviation above approximately 25\% concentration for both species. A purely data-driven model without physical constraints achieves comparable concentration accuracy (Supplementary Note~4). These results establish that coupled transient responses to variable gas inputs can be inverted into continuous, time-resolved concentration trajectories without assuming a single fixed waveform.

Beyond concentration accuracy, the constrained inversion yields a jointly consistent set of outputs whose internal states and signal reconstruction can be examined directly.

\begin{figure*}[htbp]
    \centering
    \includegraphics[width=\textwidth]{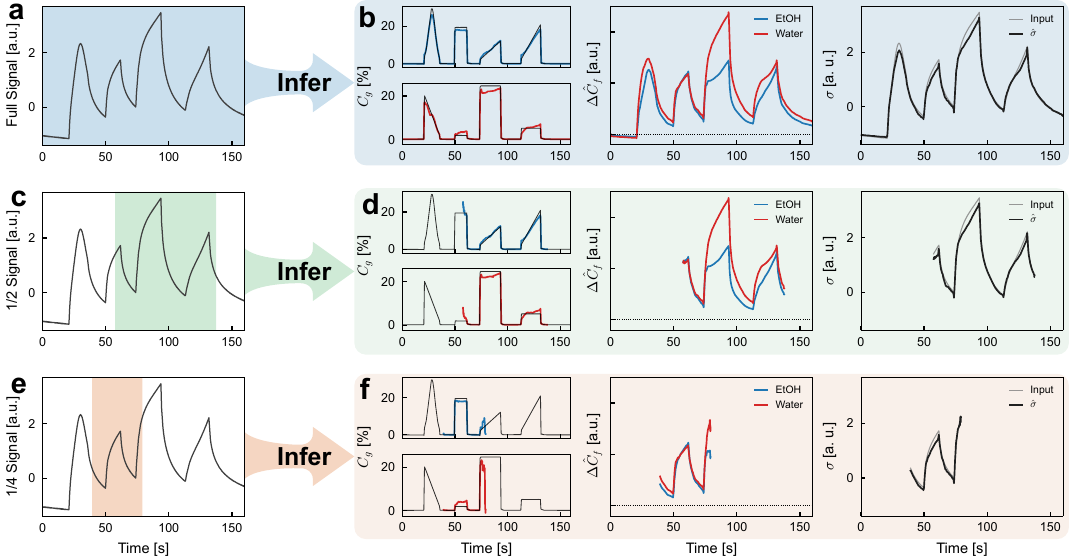}
    \caption{Joint reconstruction of gas concentrations, film-phase states, and the sensor response under variable input windows.
    (a,~c,~e)~Measured multichannel sensor signals for a representative test sequence. Shaded regions indicate the input time windows: full length~(a), half length~(c), and quarter length~(e).
    (b,~d,~f)~Reconstruction results corresponding to the input windows in (a,~c,~e), each arranged in three columns. Left: reconstructed gas-phase concentrations $\hat{C}_g$ (colored) and ground truth (black solid) for ethanol (top) and water (bottom). Center: inferred film-phase concentrations $\hat{C}_{f}$ for the corresponding gas--channel pairs. Right: measured sensor signal $\sigma$ (gray) and reconstructed $\hat{\sigma}$ (black) for a representative channel. Shorter windows produce minor distortions near temporal boundaries, but the central regions remain consistent with the full-length result.}
    \label{fig:joint_inference}
\end{figure*}

Figure~\ref{fig:joint_inference}a,b shows the joint reconstruction for a representative test sequence at full input length. The reconstructed signal $\hat{\sigma}$ closely tracks the measured input across all channels. The inferred $\hat{C}_f$ exhibits temporally delayed dynamics relative to $\hat{C}_g$, reflecting the finite sorption time scale imposed by Eq.~\eqref{eq:sorption}. Characteristic time-scale differences between species are visible in the relative dynamics of their $\hat{C}_f$ trajectories, consistent with the learned sorption parameters (Supplementary Table~S3). The learned viscoelastic parameters, in contrast, reveal a degeneracy: the modulus ratio $E_U/E_R$ converges to approximately unity across all channels (Supplementary Note~3), so the SLS model effectively reduces to an elastic response under the present conditions. This arises because the sorption time constant $\tau_s$ filters fast concentration transients before they reach the mechanical element, leaving the strain rate too low to resolve viscoelastic relaxation. The viscoelastic constraint nonetheless remains structurally important: it enforces temporal smoothness on the $C_f \rightarrow \sigma$ pathway. Its residual $\mathcal{R}_2$ is also sensitive to dynamical discontinuities that the signal-level metric alone cannot identify.

When the input window is shortened to one-half (Fig.~\ref{fig:joint_inference}c,d) or one-quarter (Fig.~\ref{fig:joint_inference}e,f) of the full sequence length, the reconstructed trajectories show minor distortions near the temporal boundaries but remain consistent in the central region. This confirms that the framework does not require a specific input duration or waveform structure.

Because $\hat{\sigma}$ is derived from the same internal state as $\hat{C}_g$ yet corresponds to a directly measured quantity, any discrepancy between $\hat{\sigma}$ and the measured $\sigma$ constitutes a computable consistency indicator that requires no external ground truth. The physical residuals $\mathcal{R}_1$ and $\mathcal{R}_2$ (Eqs.~\eqref{eq:r1}--\eqref{eq:r2}) provide a complementary diagnostic: they are sensitive to local violations of the governing dynamics even when the signal-level mismatch remains small. Together, these two layers of indicators offer both global and local anomaly detection from the inference outputs alone. The next section tests both under controlled perturbations.

\section{Reconstruction mismatch provides an intrinsic confidence readout}\label{sec:selfverify}

We now test whether the reconstructed signal and the physical residuals detect input anomalies in practice, using two classes of controlled perturbation that target different failure modes.

\subsection{Global signal perturbations detected by signal inconsistency}\label{subsec:global}

\begin{figure*}[htbp]
    \centering
    \includegraphics[width=\textwidth]{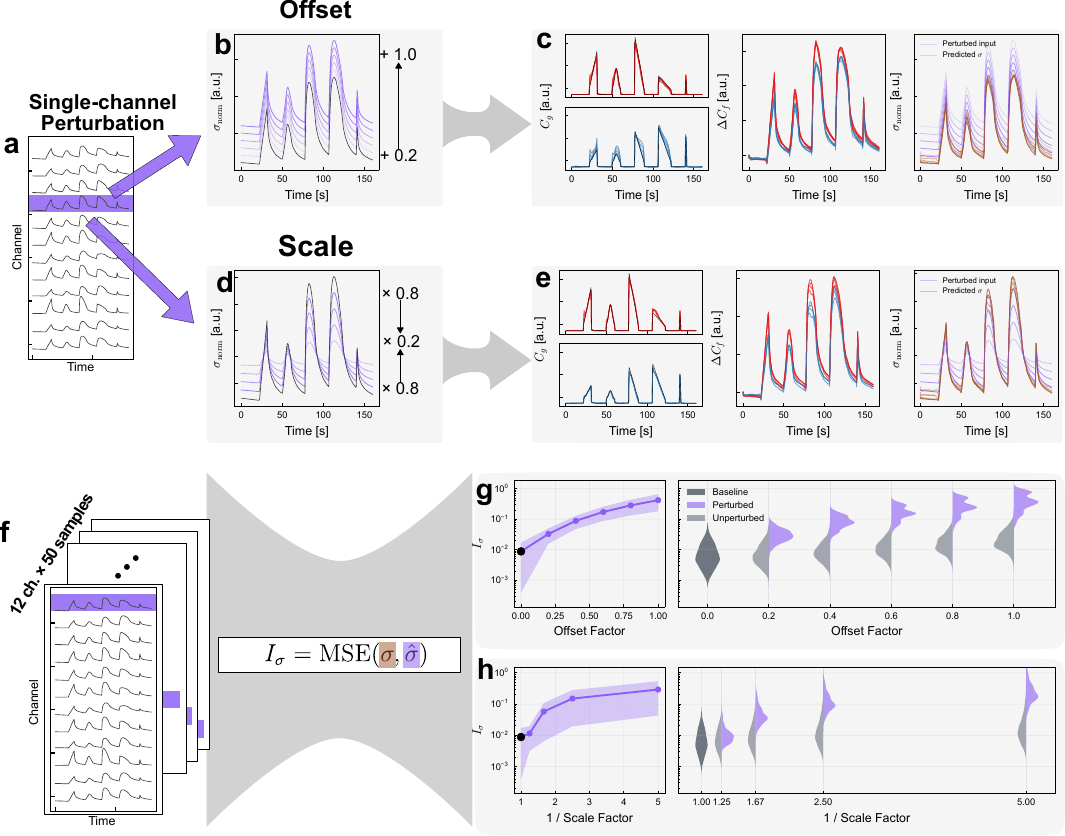}
    \caption{Self-verification under global single-channel perturbations.
    (a)~A single channel (highlighted) is randomly selected from the twelve-channel input and subjected to two perturbation types: an additive offset~(b) and a multiplicative scaling~(d).
    (b)~Offset perturbation applied to the selected channel in z-score normalized space (representative offsets up to $+1.0$; black: unperturbed).
    (c)~Model outputs under offset perturbation. Left: reconstructed gas-phase concentrations $\hat{C}_g$ (colored) and ground truth (black) for ethanol (top) and water (bottom). Center: inferred film-phase concentrations $\hat{C}_f$. Right: the reconstructed $\hat{\sigma}$ (brown) diverges from the perturbed input (purple), exposing the inconsistency.
    (d)~Scale perturbation, multiplying the selected channel by factors down to $0.2$.
    (e)~Model outputs under scale perturbation, same layout as (c).
    (f)~Signal inconsistency evaluated over the full test set (twelve channels $\times$ 50 samples): $I_{\sigma} = \mathrm{MSE}(\sigma, \hat{\sigma})$ between the (perturbed) input $\sigma$ and the reconstruction $\hat{\sigma}$, computed independently for each channel.
    (g)~Offset perturbation. Left: $I_{\sigma}$ versus offset magnitude; the line and shaded band show the mean and standard deviation across the test set, and the black dot marks the unperturbed baseline. Right: distributions of $I_{\sigma}$ at selected offset magnitudes for the baseline (dark gray), the perturbed channel (purple), and the unperturbed channels (light gray).
    (h)~Same as (g) for scale perturbation, plotted against the inverse scale factor $1/\alpha$.}
    \label{fig:perturbation}
\end{figure*}

For each test sequence, a single channel is randomly selected (Fig.~\ref{fig:perturbation}a) and perturbed in z-score normalized signal space to ensure uniform perturbation strength across channels with different raw amplitudes. Two perturbation types are applied: (i)~offset, $\sigma'_j \to \sigma'_j + \delta$ with $\delta \in [0, 1]$ (Fig.~\ref{fig:perturbation}b), and (ii)~scale, $\sigma'_j \to \alpha\,\sigma'_j$ with $\alpha \in [0.2, 1]$ (Fig.~\ref{fig:perturbation}d). These signatures correspond to common failure modes of deployed sensors. Offset shifts arise from receptor contamination, thermal drift, or electronic bias, and scale changes arise from receptor poisoning, coating degradation, or gain errors. The same signatures can equally be introduced by deliberate interference in the readout chain. The perturbed input is passed through the frozen model to obtain $\hat{C}_g$, $\hat{C}_f$, and $\hat{\sigma}$.

Figure~\ref{fig:perturbation}c,e shows that under both perturbation types, the inferred $\hat{C}_g$ and $\hat{C}_f$ exhibit only modest deviations from their unperturbed trajectories, while the reconstructed $\hat{\sigma}$ progressively diverges from the perturbed input. The model does not reproduce the corrupted signal but instead projects it toward the learned response manifold, the set of signals attainable under the training data and the governing dynamics, making the corruption visible as a mismatch. The stability of the inferred concentrations is itself desirable, but a user at deployment has no way to know whether an estimate survived the corruption. $I_{\sigma}$ supplies that evidence.

To quantify this mismatch, we define the signal inconsistency (Fig.~\ref{fig:perturbation}f), computed independently for each channel:
\begin{equation}\label{eq:isigma}
    I_{\sigma} = \frac{1}{N}\sum_{i=1}^{N} \left({\sigma_i - \hat{\sigma}_i}\right)^2,
\end{equation}
where $\sigma_i$ is the (perturbed) input, $\hat{\sigma}_i$ the reconstructed signal, and $N$ the number of time steps in the acquisition window. As shown in Fig.~\ref{fig:perturbation}g, $I_{\sigma}$ increases monotonically with offset magnitude across the full test set, growing approximately linearly. For scale perturbations (Fig.~\ref{fig:perturbation}h), a rapid initial increase is followed by gradual saturation at larger amplitudes. In both cases, the per-channel distributions confirm that the perturbed channel exhibits markedly elevated $I_{\sigma}$ relative to unperturbed channels, indicating that the metric localizes the source of inconsistency.

Ablation analysis (Supplementary Note~4) confirms that $I_{\sigma}$ operates even in a data-only baseline without explicit physical constraints, because it originates from the joint signal reconstruction objective rather than from the governing equations themselves. The physical constraints, however, tighten the learned manifold and sharpen detection sensitivity.

\subsection{Local dynamical violations detected by physical residuals}\label{subsec:local}

\begin{figure*}[htbp]
    \centering
    \includegraphics[width=\textwidth]{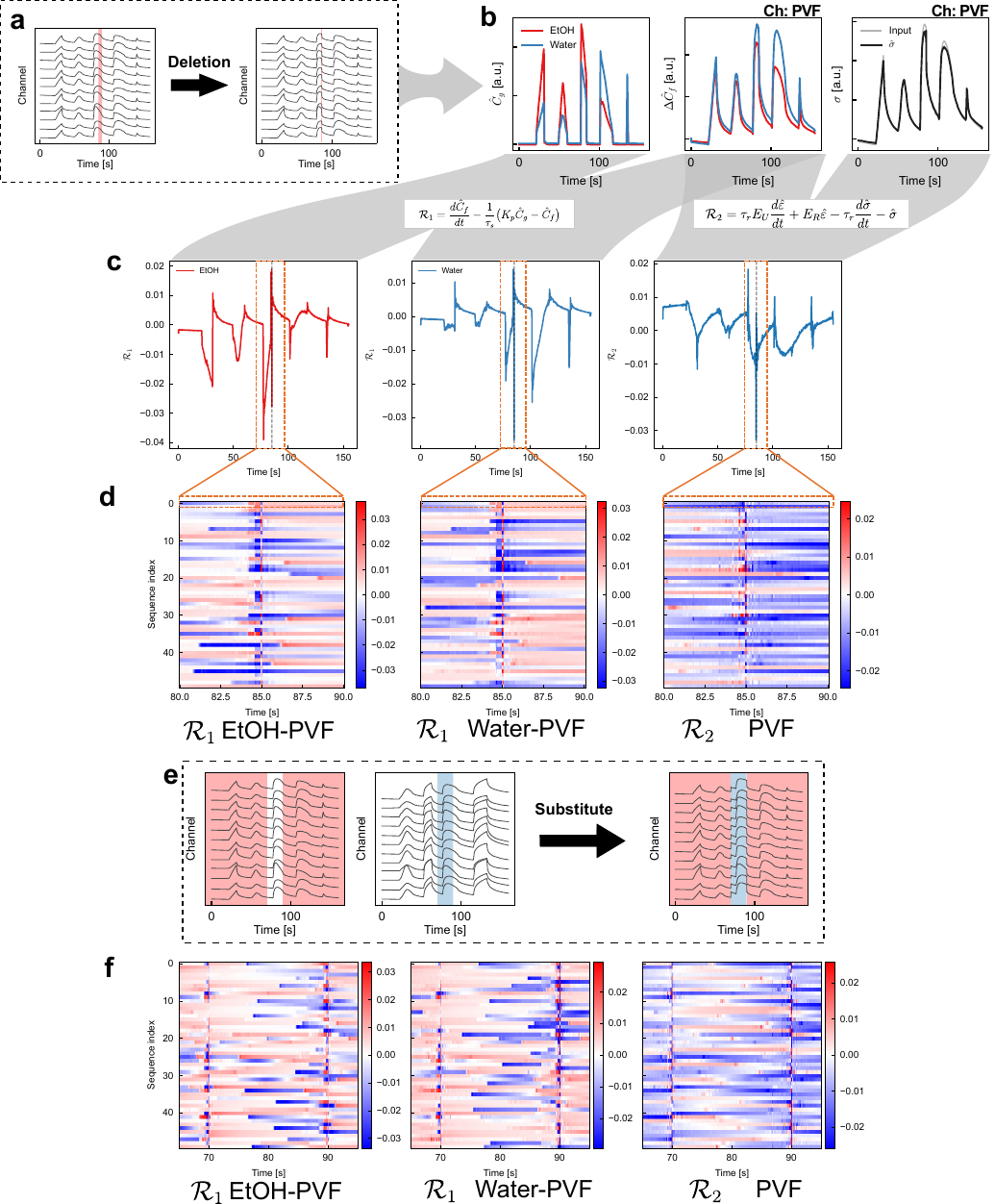}
    \caption{Detection of local dynamical discontinuities by physical residuals.
    (a)~Deletion splice: a segment is removed from a test sequence and the remaining portions are concatenated, introducing a temporal discontinuity that preserves signal statistics.
    (b)~Model outputs for the spliced input (representative channel PVF). Left: reconstructed gas-phase concentrations $\hat{C}_g$ for ethanol (red) and water (blue). Center: inferred film-phase concentrations $\hat{C}_f$. Right: measured input $\sigma$ (gray) and reconstructed $\hat{\sigma}$ (black). The reconstruction shows no visible mismatch, and $I_{\sigma}$ remains at baseline levels.
    (c)~Physical residuals for a representative sequence: the sorption residual $\mathcal{R}_1$ for ethanol (left) and water (center), and the viscoelastic residual $\mathcal{R}_2$ (right). Sharp spikes appear at the splice location (dashed box), detecting the dynamical discontinuity that $I_{\sigma}$ misses.
    (d)~Residual heatmaps across the full test set (horizontal axis: time around the splice; vertical axis: sequence index) for $\mathcal{R}_1$ (ethanol--PVF), $\mathcal{R}_1$ (water--PVF), and $\mathcal{R}_2$ (PVF). Hotspots at the splice location confirm that the detection is consistent across sequences.
    (e)~Substitution splice: the 70--90~s segment of a test sequence is replaced by the corresponding segment of another measured sequence, so that every sample of the record is genuine sensor data and the only anomalies are the dynamical breaks at the two junctions.
    (f)~Residual heatmaps across the full test set for substitution splices, same layout as (d). Hotspots concentrate at both junction positions and bracket the substituted region, while $I_{\sigma}$ remains at baseline.
    Residual panels show the representative channel PVF; detection statistics across all twelve channels are provided in Supplementary Note~4.
    }
    \label{fig:splice}
\end{figure*}

The global perturbations above alter the signal amplitude and are readily captured by $I_{\sigma}$. A different class of anomaly, however, can evade signal-level detection entirely. To demonstrate this, we construct a deletion splice, in which a temporal segment is removed from a test sequence and the remaining portions are concatenated (Fig.~\ref{fig:splice}a). Such discontinuities arise in practice when acquisition buffers overflow or when a data logger restarts during measurement. Because the signal values on either side of the splice individually lie within the training distribution, the spliced input appears statistically normal.

Figure~\ref{fig:splice}b confirms that the model reconstructs $\hat{\sigma}$ without visible mismatch, and $I_{\sigma}$ remains at baseline levels. The splice is invisible at the signal level. However, the physical residuals $\mathcal{R}_1$ and $\mathcal{R}_2$ computed from the model outputs (Fig.~\ref{fig:splice}c) exhibit sharp spikes at the splice location. The discontinuity violates the temporal continuity assumed by the governing equations (Eqs.~\eqref{eq:sorption}--\eqref{eq:sls}), and this violation is detected through the residuals even though the reconstruction itself appears normal. Residual heatmaps across the full test set (Fig.~\ref{fig:splice}d) confirm that the detection is consistent, with hotspots concentrated at the splice locations.

A stricter variant replaces the 70--90~s segment of each test sequence with the corresponding segment of another measured sequence (Fig.~\ref{fig:splice}e). Every sample of the resulting record is genuine sensor data, and because all sequences share the same exposure phase structure, the substituted segment is statistically aligned with its surroundings. The only anomalies are the dynamical breaks at the two junctions. This construction mimics the post-hoc editing of a measurement record, in which an unwanted episode is replaced by a clean one. As in the deletion case, the reconstructed $\hat{\sigma}$ tracks the input without visible mismatch, and $I_{\sigma}$ increases by less than 7\% on average across the test set, well within its baseline spread. The physical residuals, in contrast, spike at both junctions (Fig.~\ref{fig:splice}f). The two hotspots bracket the substituted region, so the residuals not only detect the edit but delimit it. Physical consistency thus provides a tamper-evident layer for sensor records that signal statistics alone cannot supply.

\subsection{Complementary roles of signal-level and physics-level diagnostics}\label{subsec:complementary}

These experiments reveal that $I_{\sigma}$ and $\mathcal{R}_1$/$\mathcal{R}_2$ respond to fundamentally different classes of anomaly. Under offset and scale perturbations, $I_{\sigma}$ rises monotonically while $\mathcal{R}_1$ and $\mathcal{R}_2$ remain comparatively stable (Supplementary Note~4), because the model projects the corrupted input onto the learned response manifold without distorting its internal dynamics. Under both splice variants, the situation is reversed. $I_{\sigma}$ is blind to the discontinuity, but $\mathcal{R}_1$ and $\mathcal{R}_2$ detect the local dynamical violation directly. Together, they form a layered diagnostic, in which $I_{\sigma}$ flags global distributional shifts in the input signal, while $\mathcal{R}_1$ and $\mathcal{R}_2$ flag local violations of the governing dynamics.

\FloatBarrier
\section{Discussion}\label{sec:conclusion}

We have presented a constrained dynamical inversion framework that reconstructs time-varying multicomponent gas concentrations from multichannel sensor signals without requiring a single fixed waveform, a predefined input duration, or equilibrium feature extraction. The framework recovers continuous concentration trajectories under both synchronous and asynchronous mixing conditions and remains stable across variable input window lengths. By requiring the model to jointly reconstruct the measured sensor response $\hat{\sigma}$ from the same internal state that produces the inferred concentrations $\hat{C}_g$, the framework further turns self-verification into an intrinsic by-product of inference. The signal inconsistency $I_{\sigma}$ detects global distributional shifts in the input, while the physical residuals $\mathcal{R}_1$ and $\mathcal{R}_2$ detect local dynamical violations invisible at the signal level. Neither requires access to ground-truth concentrations, and their combination covers a broader range of failure modes than either alone. Full-composition inference also underpins the self-verification itself. The reconstructed signal can only be checked because every contribution to it is claimed by an inferred component. The framework thus turns transients into information and interferents into estimated states.

The perturbation experiments also suggest relevance to measurement-record integrity. The diagnostics respond to classes of physical inconsistency rather than to their causes. An offset in one channel may originate from receptor contamination or from a bias introduced into the readout chain. A dynamical discontinuity may originate from a logger restart or from post-hoc editing of a stored record. In each case the framework reports the inconsistency and localizes it, while attributing the cause requires external context. For maintenance, $I_{\sigma}$ acts as an indicator of drift and sensitivity loss that identifies the affected channel. For record integrity, the physical residuals detected the tested deletion and substitution edits, including a case in which every sample was genuine data. This does not constitute a complete data-integrity or authentication system, but it adds a physics-based check without additional hardware.

Several limitations should be noted. Although the framework is parametric in the number of gas species and sensor channels, the current experiments are restricted to binary water--ethanol mixtures. Static regression on the same MSS platform has achieved ternary water--methanol--ethanol quantification~\cite{shiba2018a}, suggesting that extension to higher-order mixtures is feasible in principle. A natural open question is how many gas components can be jointly resolved for a given channel count under constrained inversion, and how this limit depends on receptor diversity and the information contributed by the physical constraints. As noted in Section~\ref{sec:reconstruction}, the SLS model reduces to an elastic response under the present conditions ($E_U/E_R \approx 1$), though its structural role in enforcing temporal consistency persists. Additionally, the framework assumes that the governing equations adequately describe the sensing dynamics within the training distribution. If the true dynamics depart from this assumption, the physical residuals may increase without a corresponding change in $I_{\sigma}$, a failure mode that should be addressed through more expressive constitutive models or online parameter adaptation in future work. The integrity diagnostics have boundaries of their own. A forged record that satisfies the governing dynamics, such as a replayed segment of previously measured clean data, produces no inconsistency and cannot be detected. Manipulations acting before the gas reaches the sensor, such as sample dilution, yield genuinely consistent signals and likewise fall outside the scope of these diagnostics.

More broadly, this study reframes chemical sensing from static pattern recognition to constrained state inference. Rather than returning point estimates with no internal quality criterion, the framework produces concentration trajectories together with computable evidence of whether those trajectories are internally consistent. Future directions include extension to higher-order mixtures, integration of $I_{\sigma}$ and residual thresholds into automated decision protocols, and application to single-sample measurement scenarios where repeated calibration is impractical.

\section{Methods}\label{sec:methods}

\subsection{Sensor array and experimental measurement}\label{subsec:sensor}

Bare MSS chips were coated with twelve chemically distinct polymer receptor layers: poly(vinylidene fluoride) (PVF), polystyrene (PS), polycaprolactone (PCL), poly(methyl methacrylate) (PMMA), cellulose acetate butyrate (CAB), Tenax TA (TENAX), poly(4-methylstyrene) (P4MS), polysulfone (PSF), poly(4-vinylphenol) (PVP), poly(vinyl chloride) (PVC), poly(methyl vinyl ether-\textit{alt}-maleic anhydride) (PMVE), and poly(bisphenol A carbonate) (PBA). All polymers were purchased from Sigma-Aldrich Inc., except Tenax TA (GL Sciences), and used as received. Each polymer was dissolved in \textit{N,N}-dimethylformamide (DMF; FUJIFILM Wako Pure Chemical, Guaranteed Reagent) or 1,1,2,2-tetrachloroethane (TCE; Nacalai Tesque) and deposited onto the chips with an inkjet spotter (LaboJet-500SP, MICROJET Co.\ Ltd.) fitted with a nozzle (IJHBS-300, MICROJET Co.\ Ltd.), followed by solvent evaporation at 60$^{\circ}$C and drying at room temperature for 24~h. Three four-channel chips were mounted in a sensor cell for gas exposure. Sensor signals were acquired at 100~Hz and downsampled to 20~Hz for analysis.

\subsection{Gas-delivery protocols}\label{subsec:gasdelivery}

Dynamic binary vapor mixtures of water and ethanol were generated using three mass flow controllers (MFCs) with N$_2$ as the carrier gas. MFC-1 and MFC-2 passed N$_2$ through bubblers containing liquid ethanol (FUJIFILM Wako Pure Chemical, Guaranteed Reagent) and water (ultrapure, Milli-Q; Merck Millipore), respectively. MFC-3 provided dilution N$_2$. The mixture composition was controlled by dynamically regulating MFC flow rates according to four waveform patterns (rectangular, sawtooth, reverse sawtooth, and Gaussian), selected to approximate common exposure profiles.

Each measurement sequence lasts 240~s: an initial purge (20~s), an active injection period (120~s) containing 4--5 waveform segments of 10--20~s interspersed with purge intervals, and a recovery purge (100~s). Data were collected under two regimes: synchronous mixing, where both MFCs follow the same waveform, and asynchronous mixing, where each MFC follows an independently selected waveform. The training set comprises 150 sequences (80 synchronous, 70 asynchronous) and the test set 50 sequences (20 synchronous, 30 asynchronous). Each sequence is truncated to 160~s (3200 time steps) for model input. Further details on data generation and dataset composition are provided in Supplementary Note~2.

\subsection{Physics-based forward model}\label{subsec:forwardmodel}

The forward dynamics consist of two coupled stages. Sorption kinetics (Eq.~\eqref{eq:sorption}) governs mass transport from the gas phase into the receptor film, parameterized by a sorption time constant $\tau_s$ and partition coefficient $K_p$ for each gas--sensor pair. The mechanical response follows a standard linear solid model (Eq.~\eqref{eq:sls}), coupled to the film-phase concentration through the small-strain relation $\varepsilon = (v/3)\,C_f$. For an array of $n_{\mathrm{sensors}}$ sensors exposed to $n_{\mathrm{gases}}$ gas species, each gas--sensor pair has independent sorption parameters $(\tau_{s,ij}, K_{p,ij})$, and each sensor has independent viscoelastic parameters $(\tau_{r,j}, E_{U,j}, E_{R,j})$. All physical parameters are treated as learnable quantities during training. The complete multi-gas, multi-sensor formulation and the normalization scheme used for numerical stability are given in Supplementary Note~3.

\subsection{Neural network architecture}\label{subsec:architecture}

The model consists of a non-causal temporal convolutional network (TCN) encoder with 11 layers of dilated convolutions (dilation factors $2^0, 2^1, \ldots, 2^{10}$, kernel size 3, 64 output channels), yielding a receptive field of 4095 time steps that covers the full input length. The TCN output is decoded by a shared MLP (two hidden layers, 512 units each), which connects to three prediction heads (one hidden layer, 128 units each) for $\hat{C}_g$, $\hat{C}_f$, and $\hat{\sigma}$. Dropout ($p = 0.1$) is applied after each hidden layer. The complete hyperparameter table is provided in Supplementary Table~S1.

\subsection{Training objectives and residual constraints}\label{subsec:training}

The primary training objective is a weighted sum of the signal reconstruction loss $\mathcal{L}_V = \mathrm{MSE}(V, \hat{V})$, where $V$ denotes the raw sensor output (equivalent to $\sigma$ after absorbing the stress-to-electrical conversion factor), and the concentration prediction loss $\mathcal{L}_{C_g} = \mathrm{MSE}(C_g, \hat{C}_g)$. The physical residuals $\mathcal{R}_1$ (Eq.~\eqref{eq:r1}) and $\mathcal{R}_2$ (Eq.~\eqref{eq:r2}), together with the constraint $E_U/E_R \geq 1$, are enforced as soft constraints through the modified differential multiplier method (MDMM)~\cite{platt,almanstotter2025}, which automatically balances constraint penalties against the data loss during training. Time derivatives in the residuals are approximated by central finite differences. The model is optimized with Adan~\cite{xie2024a} using a ReduceLROnPlateau learning rate schedule (initial rate $10^{-3}$, decay factor 0.5, patience 200, minimum rate $10^{-5}$). Training is terminated by early stopping with a patience of 800 epochs. All input signals are z-score normalized using training-set statistics, with the same normalization applied to test data without refitting. Training convergence across five independent random seeds is shown in Supplementary Fig.~S2.

\subsection{Perturbation experiments}\label{subsec:perturbation}

Two classes of perturbation are applied to test-set sequences after training, with all network parameters frozen. These perturbations are designed to mimic common real-world sensor failure modes. For global perturbations (Section~\ref{subsec:global}), a single channel is randomly selected and modified in z-score normalized signal space. Offset perturbation adds a constant $\delta \in [0, 1]$ to the channel, simulating baseline drift, and scale perturbation multiplies the channel by a factor $\alpha \in [0.2, 1]$, simulating sensitivity degradation. For splice perturbations (Section~\ref{subsec:local}), two variants are applied. In the deletion variant, a temporal segment is removed from a test sequence and the remaining portions are concatenated, simulating data acquisition interruptions while preserving signal-level statistics. In the substitution variant, the 70--90~s segment of each test sequence is replaced by the corresponding segment of the next sequence in the test set, so that every sample of the record is genuine measured data and only the two junctions violate the governing dynamics, simulating post-hoc editing of a stored record.
In all cases, the perturbed input is passed through the frozen model to obtain $\hat{C}_g$, $\hat{C}_f$, and $\hat{\sigma}$, from which $I_{\sigma}$, $\mathcal{R}_1$, and $\mathcal{R}_2$ are computed.

\subsection{Evaluation metrics}\label{subsec:metrics}

Concentration reconstruction accuracy is evaluated by the root-mean-square error (RMSE) and coefficient of determination ($R^2$) between reconstructed and ground-truth concentration trajectories across all test sequences. The signal inconsistency $I_{\sigma}$ (Eq.~\eqref{eq:isigma}) quantifies the mean squared error between the measured input and the reconstructed signal for each channel. Physical residuals $\mathcal{R}_1$ and $\mathcal{R}_2$ (Eqs.~\eqref{eq:r1}--\eqref{eq:r2}) are evaluated pointwise along each trajectory. Per-channel RMSE, $R^2$, and cumulative error distributions are reported in Supplementary Note~5.

\backmatter

\bmhead{Data availability}
The data that support the findings of this study are available from the corresponding author upon reasonable request.

\bmhead{Code availability}
The source code for data processing, model training, inference, and figure generation will be made publicly available on GitHub upon publication.

\bmhead{Acknowledgements}
Y.Z. thanks the NIMS Junior Research Program, NIMS, Japan. Y.Z. is grateful for the support by the Pioneering Research Initiated by the Next Generation (SPRING) program, JST, MEXT, Japan (No.\ JPMJSP2124).

\bmhead{Author contributions}
Yingcheng Zhou: Conceptualization, Methodology, Software, Formal analysis, Investigation, Visualization, Writing -- original draft, Funding acquisition. Kosuke Minami: Writing -- review \& editing, Supervision. Genki Yoshikawa: Writing -- review \& editing, Supervision.

\bmhead{Competing interests}
The authors declare no competing interests.

\bibliography{main}

\begin{thebibliography}{10}
\expandafter\ifx\csname url\endcsname\relax
  \def\url#1{\burl{#1}}\fi
\expandafter\ifx\csname urlprefix\endcsname\relax\def\urlprefix{URL }\fi
\providecommand{\bibinfo}[2]{#2}
\providecommand{\eprint}[2][]{\url{#2}}
\providecommand{\doi}[1]{\url{https://doi.org/#1}}
\bibcommenthead

\bibitem{zong2024}
\bibinfo{author}{Zong, B.} \emph{et~al.}
\newblock \bibinfo{title}{Smart {{Gas Sensors}}: {{Recent Developments}} and
  {{Future Prospective}}}.
\newblock \emph{\bibinfo{journal}{Nano-Micro Lett.}}
  \textbf{\bibinfo{volume}{17}}, \bibinfo{pages}{54} (\bibinfo{year}{2024}).

\bibitem{maji2025}
\bibinfo{author}{Maji, S.} \emph{et~al.}
\newblock \bibinfo{title}{Molecular {{Nanoarchitectonic Sensing Layer}} for
  {{Analysis}} of {{Volatile Fatty Acids}} in {{Bioreactor Headspaces Using}} a
  {{Nanomechanical Sensor}}}.
\newblock \emph{\bibinfo{journal}{Adv. Mater. Technol.}}
  \textbf{\bibinfo{volume}{10}}, \bibinfo{pages}{e00294}
  (\bibinfo{year}{2025}).

\bibitem{mor2025}
\bibinfo{author}{Mor, S.} \emph{et~al.}
\newblock \bibinfo{title}{Current {{Opportunities}} and {{Trends}} in the {{Gas
  Sensor Market}}: {{A Focus}} on e-{{Noses}} and {{Their Applications}} in
  {{Food Industry}}}.
\newblock \emph{\bibinfo{journal}{Chemosensors}} \textbf{\bibinfo{volume}{13}},
  \bibinfo{pages}{181} (\bibinfo{year}{2025}).

\bibitem{loizeau2013}
\bibinfo{author}{Loizeau, F.} \emph{et~al.}
  \emph{\bibinfo{title}{Piezoresistive membrane-type surface stress sensor
  arranged in arrays for cancer diagnosis through breath analysis}}.
\newblock In \emph{\bibinfo{booktitle}{2013 {{IEEE}} 26th {{Int}}. {{Conf}}.
  {{Micro Electro Mech}}. {{Syst}}. {{MEMS}}}}, \bibinfo{pages}{621--624}
  (\bibinfo{publisher}{IEEE}, \bibinfo{address}{Taipei, Taiwan},
  \bibinfo{year}{2013}).

\bibitem{saeki2024b}
\bibinfo{author}{Saeki, Y.} \emph{et~al.}
\newblock \bibinfo{title}{Lung cancer detection in perioperative patients'
  exhaled breath with nanomechanical sensor array}.
\newblock \emph{\bibinfo{journal}{Lung Cancer}} \textbf{\bibinfo{volume}{190}},
  \bibinfo{pages}{107514} (\bibinfo{year}{2024}).

\bibitem{devkota2018}
\bibinfo{author}{Devkota, J.} \emph{et~al.}
\newblock \bibinfo{title}{Zeolitic imidazolate framework-coated acoustic
  sensors for room temperature detection of carbon dioxide and methane}.
\newblock \emph{\bibinfo{journal}{Nanoscale}} \textbf{\bibinfo{volume}{10}},
  \bibinfo{pages}{8075--8087} (\bibinfo{year}{2018}).

\bibitem{matatagui2019}
\bibinfo{author}{Matatagui, D.}, \bibinfo{author}{Bahos, F.~A.},
  \bibinfo{author}{Gr{\`a}cia, I.} \& \bibinfo{author}{Horrillo, M. d.~C.}
\newblock \bibinfo{title}{Portable {{Low-Cost Electronic Nose Based}} on
  {{Surface Acoustic Wave Sensors}} for the {{Detection}} of {{BTX Vapors}} in
  {{Air}}}.
\newblock \emph{\bibinfo{journal}{Sensors}} \textbf{\bibinfo{volume}{19}},
  \bibinfo{pages}{5406} (\bibinfo{year}{2019}).

\bibitem{oon-pitipongsa2026}
\bibinfo{author}{{Oon-pitipongsa}, T.} \emph{et~al.}
\newblock \bibinfo{title}{Convolutional {{Neural Networks}} on {{Correlation}}
  between {{GC}}-{{MS Molecular Data}} and {{QCM Gas-Sensing Data}}}.
\newblock \emph{\bibinfo{journal}{ACS Sens.}} \textbf{\bibinfo{volume}{11}},
  \bibinfo{pages}{1152--1161} (\bibinfo{year}{2026}).

\bibitem{malhotra2023}
\bibinfo{author}{Malhotra, J.~S.}, \bibinfo{author}{Kubus, M.},
  \bibinfo{author}{Pedersen, K.~S.}, \bibinfo{author}{Andersen, S.~I.} \&
  \bibinfo{author}{Sundberg, J.}
\newblock \bibinfo{title}{Room-{{Temperature Monitoring}} of {{CH4}} and {{CO2
  Using}} a {{Metal}}--{{Organic Framework-Based QCM Sensor Showing Inherent
  Analyte Discrimination}}}.
\newblock \emph{\bibinfo{journal}{ACS Sens.}} \textbf{\bibinfo{volume}{8}},
  \bibinfo{pages}{3478--3486} (\bibinfo{year}{2023}).

\bibitem{minami2025}
\bibinfo{author}{Minami, K.} \& \bibinfo{author}{Yoshikawa, G.}
\newblock \bibinfo{title}{An {{Analytical Model}} of {{Sorption-Induced Static
  Mode Nanomechanical Sensing}} for {{Multicomponent Analytes}}}.
\newblock \emph{\bibinfo{journal}{Anal. Chem.}} \textbf{\bibinfo{volume}{97}},
  \bibinfo{pages}{19306--19312} (\bibinfo{year}{2025}).

\bibitem{minami2024}
\bibinfo{author}{Minami, K.}, \bibinfo{author}{Zhou, Y.},
  \bibinfo{author}{Imamura, G.}, \bibinfo{author}{Shiba, K.} \&
  \bibinfo{author}{Yoshikawa, G.}
\newblock \bibinfo{title}{Sorption {{Kinetic Parameters}} from {{Nanomechanical
  Sensing}} for {{Discrimination}} of 2-{{Nonenal}} from {{Saturated
  Aldehydes}}}.
\newblock \emph{\bibinfo{journal}{ACS Sens.}} \textbf{\bibinfo{volume}{9}},
  \bibinfo{pages}{689--698} (\bibinfo{year}{2024}).

\bibitem{imamura2018c}
\bibinfo{author}{Imamura, G.}, \bibinfo{author}{Shiba, K.},
  \bibinfo{author}{Yoshikawa, G.} \& \bibinfo{author}{Washio, T.}
\newblock \bibinfo{title}{Analysis of nanomechanical sensing signals; physical
  parameter estimation for gas identification}.
\newblock \emph{\bibinfo{journal}{AIP Adv.}} \textbf{\bibinfo{volume}{8}},
  \bibinfo{pages}{075007} (\bibinfo{year}{2018}).

\bibitem{minami2021a}
\bibinfo{author}{Minami, K.}, \bibinfo{author}{Shiba, K.} \&
  \bibinfo{author}{Yoshikawa, G.}
\newblock \bibinfo{title}{Sorption-induced static mode nanomechanical sensing
  with viscoelastic receptor layers for multistep injection-purge cycles}.
\newblock \emph{\bibinfo{journal}{J. Appl. Phys.}}
  \textbf{\bibinfo{volume}{129}}, \bibinfo{pages}{124503}
  (\bibinfo{year}{2021}).

\bibitem{shiba2017a}
\bibinfo{author}{Shiba, K.}, \bibinfo{author}{Tamura, R.},
  \bibinfo{author}{Imamura, G.} \& \bibinfo{author}{Yoshikawa, G.}
\newblock \bibinfo{title}{Data-driven nanomechanical sensing: Specific
  information extraction from a complex system}.
\newblock \emph{\bibinfo{journal}{Sci. Rep.}} \textbf{\bibinfo{volume}{7}},
  \bibinfo{pages}{3661} (\bibinfo{year}{2017}).

\bibitem{liu2023f}
\bibinfo{author}{Liu, T.} \emph{et~al.}
\newblock \bibinfo{title}{Review on {{Algorithm Design}} in {{Electronic
  Noses}}: {{Challenges}}, {{Status}}, and {{Trends}}}.
\newblock \emph{\bibinfo{journal}{Intell. Comput.}}
  \textbf{\bibinfo{volume}{2}}, \bibinfo{pages}{0012} (\bibinfo{year}{2023}).

\bibitem{yan2015}
\bibinfo{author}{Yan, J.} \emph{et~al.}
\newblock \bibinfo{title}{Electronic {{Nose Feature Extraction Methods}}: {{A
  Review}}}.
\newblock \emph{\bibinfo{journal}{Sensors}} \textbf{\bibinfo{volume}{15}},
  \bibinfo{pages}{27804--27831} (\bibinfo{year}{2015}).

\bibitem{imamura2019a}
\bibinfo{author}{Imamura, G.}, \bibinfo{author}{Shiba, K.},
  \bibinfo{author}{Yoshikawa, G.} \& \bibinfo{author}{Washio, T.}
\newblock \bibinfo{title}{Free-hand gas identification based on transfer
  function ratios without gas flow control}.
\newblock \emph{\bibinfo{journal}{Sci. Rep.}} \textbf{\bibinfo{volume}{9}},
  \bibinfo{pages}{9768} (\bibinfo{year}{2019}).

\bibitem{fonollosa2015}
\bibinfo{author}{Fonollosa, J.}, \bibinfo{author}{Sheik, S.},
  \bibinfo{author}{Huerta, R.} \& \bibinfo{author}{Marco, S.}
\newblock \bibinfo{title}{Reservoir computing compensates slow response of
  chemosensor arrays exposed to fast varying gas concentrations in continuous
  monitoring}.
\newblock \emph{\bibinfo{journal}{Sens. Actuators B Chem.}}
  \textbf{\bibinfo{volume}{215}}, \bibinfo{pages}{618--629}
  (\bibinfo{year}{2015}).

\bibitem{feng2025a}
\bibinfo{author}{Feng, M.-Q.}, \bibinfo{author}{Minami, K.},
  \bibinfo{author}{Zhou, Y.} \& \bibinfo{author}{Yoshikawa, G.}
\newblock \bibinfo{title}{Analytical modeling and decoupling of humidity
  effects in nanomechanical sensing based on sorption kinetics and viscoelastic
  stress relaxation}.
\newblock \emph{\bibinfo{journal}{Phys. Rev. E}}
  \textbf{\bibinfo{volume}{111}}, \bibinfo{pages}{065407}
  (\bibinfo{year}{2025}).

\bibitem{karniadakis2021}
\bibinfo{author}{Karniadakis, G.~E.} \emph{et~al.}
\newblock \bibinfo{title}{Physics-informed machine learning}.
\newblock \emph{\bibinfo{journal}{Nat. Rev. Phys.}}
  \textbf{\bibinfo{volume}{3}}, \bibinfo{pages}{422--440}
  (\bibinfo{year}{2021}).

\bibitem{heinrich2009}
\bibinfo{author}{Heinrich, S.~M.}, \bibinfo{author}{Wenzel, M.~J.},
  \bibinfo{author}{Josse, F.} \& \bibinfo{author}{Dufour, I.}
\newblock \bibinfo{title}{An analytical model for transient deformation of
  viscoelastically coated beams: {{Applications}} to static-mode
  microcantilever chemical sensors}.
\newblock \emph{\bibinfo{journal}{J. Appl. Phys.}}
  \textbf{\bibinfo{volume}{105}}, \bibinfo{pages}{124903}
  (\bibinfo{year}{2009}).

\bibitem{wenzel2008b}
\bibinfo{author}{Wenzel, M.~J.}, \bibinfo{author}{Josse, F.},
  \bibinfo{author}{Heinrich, S.~M.}, \bibinfo{author}{Yaz, E.} \&
  \bibinfo{author}{Datskos, P.~G.}
\newblock \bibinfo{title}{Sorption-induced static bending of microcantilevers
  coated with viscoelastic material}.
\newblock \emph{\bibinfo{journal}{J. Appl. Phys.}}
  \textbf{\bibinfo{volume}{103}}, \bibinfo{pages}{064913}
  (\bibinfo{year}{2008}).

\bibitem{almanstotter2025}
\bibinfo{author}{Almanst{\"o}tter, M.}, \bibinfo{author}{Vetter, R.} \&
  \bibinfo{author}{Iber, D.}
\newblock \bibinfo{title}{{{PINNverse}}: {{Accurate}} parameter estimation in
  differential equations from noisy data with constrained physics-informed
  neural networks} (\bibinfo{year}{2025}).
\newblock
  \bibinfo{eprint}{{\href{https://arxiv.org/abs/2504.05248}{{arXiv:2504.05248}}}}.

\bibitem{platt}
\bibinfo{author}{Platt, J.~C.} \& \bibinfo{author}{Barr, A.~H.}
\newblock \bibinfo{title}{ in \textit{Constrained {{Differential
  Optimization}}}} In \bibinfo{editor}{Anderson, D.~Z.} (ed.)
  \emph{\bibinfo{booktitle}{Neural {{Information Processing Systems}}}}
  \bibinfo{pages}{612--621} (\bibinfo{publisher}{American Institute of
  Physics}, \bibinfo{address}{New York, NY}, \bibinfo{year}{1988}).

\bibitem{shiba2018a}
\bibinfo{author}{Shiba, K.} \emph{et~al.}
\newblock \bibinfo{title}{Functional {{Nanoparticles-Coated Nanomechanical
  Sensor Arrays}} for {{Machine Learning-Based Quantitative Odor Analysis}}}.
\newblock \emph{\bibinfo{journal}{ACS Sens.}} \textbf{\bibinfo{volume}{3}},
  \bibinfo{pages}{1592--1600} (\bibinfo{year}{2018}).

\bibitem{xie2024a}
\bibinfo{author}{Xie, X.}, \bibinfo{author}{Zhou, P.}, \bibinfo{author}{Li,
  H.}, \bibinfo{author}{Lin, Z.} \& \bibinfo{author}{Yan, S.}
\newblock \bibinfo{title}{Adan: {{Adaptive Nesterov Momentum Algorithm}} for
  {{Faster Optimizing Deep Models}}}.
\newblock \emph{\bibinfo{journal}{IEEE Trans. Pattern Anal. Mach. Intell.}}
  \textbf{\bibinfo{volume}{46}}, \bibinfo{pages}{9508--9520}
  (\bibinfo{year}{2024}).

\end{thebibliography}

\clearpage


\section*{Supplementary material}

## CONTENTS

## Supplementary Note 1. NEURAL ARCHITECTURE AND TRAINING

### 1.1. Architecture

A non-causal temporal convolutional network (TCN) with 64 output channels encodes the multi-channel time-series input[1]. The number of layers and kernel size are chosen so that the receptive field covers the full sequence length. The TCN output is decoded by a shared multi-layer perceptron (MLP) with 512 hidden units, which connects to three prediction heads: gas-phase concentration $\hat{C}_{g,i}$, film-phase concentration $\hat{C}_{f,ij}$, and stress $\hat{\sigma}_j$. The architecture is illustrated in Fig. S1.

### 1.2. Physics-informed constraints

The governing differential equations are enforced as residual constraints during training. The sorption residual $\mathscr{R}_1 \equiv \mathscr{R}_{\text{sorp}}$ and viscoelastic residual $\mathscr{R}_2 \equiv \mathscr{R}_{\text{ve}}$ are

$$\mathscr{R}_{\text{sorp},ij} = \frac{dC'_{f,ij}}{dt} - \left(A_{ij}C'_{g,i} - B_{ij}C'_{f,ij} + C_{ij}\right) \tag{S1}$$

$$\mathscr{R}_{\text{ve},j} = D_j\frac{d\varepsilon'_j}{dt} + E_j\varepsilon'_j - \left(F_j\frac{d\sigma'_j}{dt} + G_j\sigma'_j + H_j\right) \tag{S2}$$

Time derivatives are approximated by the central difference scheme:

$$\frac{dX'}{dt} = \frac{X'(t+\Delta t) - X'(t-\Delta t)}{2\Delta t} \tag{S3}$$

where $\Delta t$ is the sampling interval.

An additional constraint enforces the physical requirement that the unrelaxed modulus exceeds the relaxed modulus:

$$\frac{E_{U,j}}{E_{R,j}} \geq 1 \quad \forall j \tag{S4}$$

### 1.3. Loss function

The data loss is defined as the weighted sum of signal reconstruction and concentration prediction losses:

$$\mathscr{L}_{\text{data}} = \omega_V\mathscr{L}_V + \omega_{C_g}\mathscr{L}_{C_g} \tag{S5}$$

where $\mathscr{L}_V = \text{MSE}(V, \hat{V})$, $\mathscr{L}_{C_g} = \text{MSE}(C_g, \hat{C}_g)$, and $\omega_V$, $\omega_{C_g}$ are the respective weights.

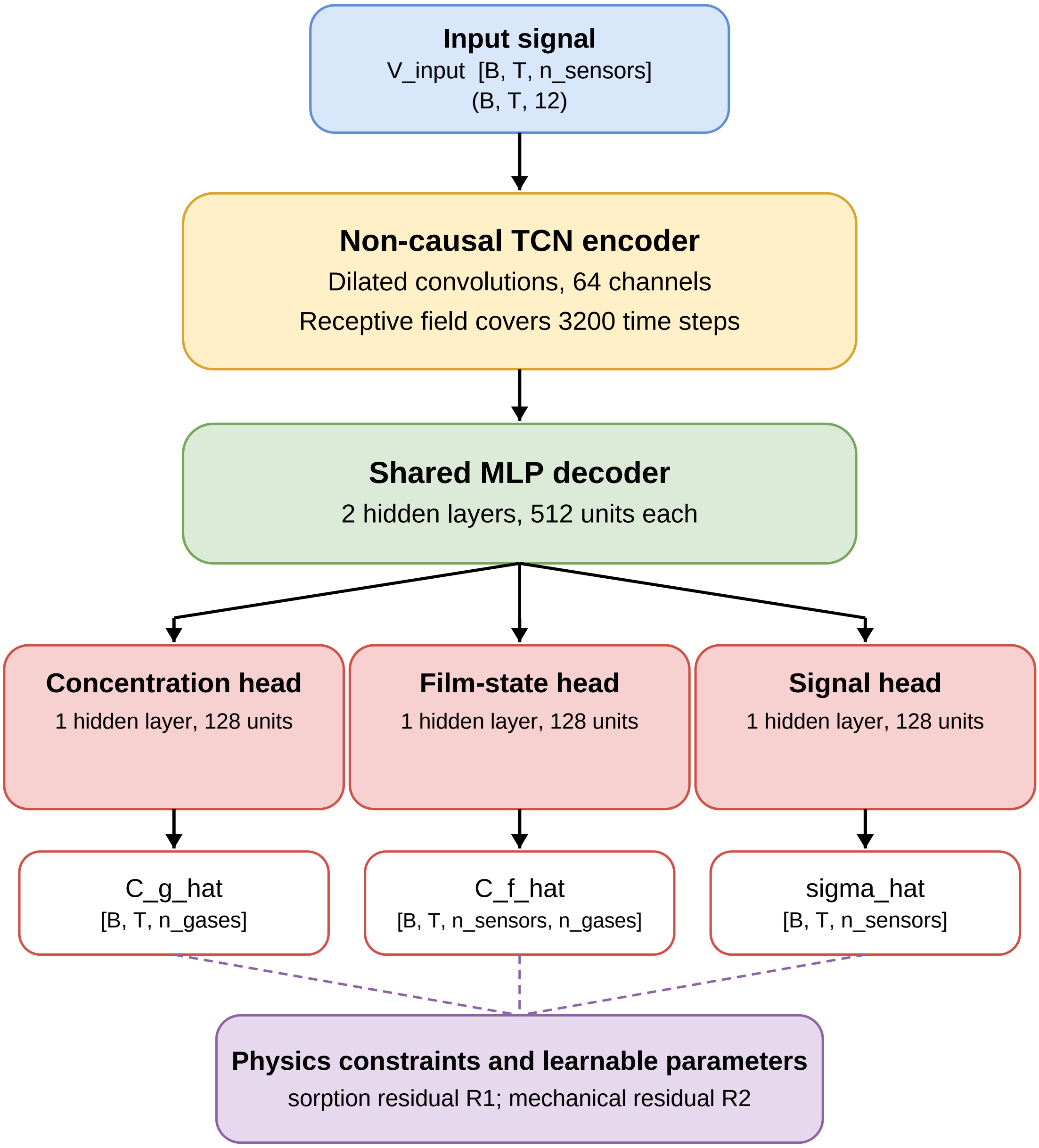


FIG. S1. Architecture of the physics-informed temporal convolutional network. The TCN encoder processes multi-channel sensor signals, and a shared MLP decoder connects to three output heads: $\hat{C}_g$, $\hat{C}_f$, and $\hat{\sigma}$.

### 1.4. Optimization

The modified differential multiplier method (MDMM) balances the data loss against the physical constraints during training. Compared to fixed-weight methods, MDMM automatically adjusts the

constraint penalty and converges to a stable equilibrium on the Pareto front[2].

The Adan optimizer[3] is used with Eqs. S1, S2, and S4 as MDMM constraints, while Eq. S5 serves as the primary objective. The learning rate is scheduled with ReduceLROnPlateau (initial rate $10^{-3}$, decay factor 0.5, patience 200, minimum rate $10^{-5}$). Training is terminated by early stopping with a patience of 800 epochs.

### 1.5. Hyperparameters

Table S1 summarizes the network architecture and training hyperparameters. The TCN uses 11 layers of non-causal dilated convolutions with exponentially increasing dilation factors ($2^0, 2^1, \dots, 2^{10}$), yielding a receptive field of 4095 time steps—larger than the input sequence length of 3200 time steps (160 s at 20 Hz). Each temporal block consists of two weight-normalized Conv1d layers with ReLU activation and residual connections. The shared MLP decoder has two hidden layers of 512 units each, and each prediction head has a single hidden layer of 128 units. Dropout ($p = 0.1$) is applied after each hidden layer throughout the network. The data loss weights are $\omega_{C_g} = 10$ and $\omega_V = 1$, reflecting the priority of concentration reconstruction. The MDMM damping parameter is set to 10 for all constraints. Input signals are z-score normalized using training-set statistics.

### 1.6. Training convergence

Figure S2 shows the training and validation loss curves for the full Physics model across five independent runs with different random seeds. All runs converge within 2000 epochs, with the total loss, concentration loss $L_C$, and signal reconstruction loss $L_V$ reaching stable plateaus. The physics residuals $\mathscr{R}_1^2$ and $\mathscr{R}_2^2$ decrease rapidly in the early epochs as the MDMM multipliers adjust, and stabilize at values several orders of magnitude below the data loss. The narrow spread across seeds indicates that training is robust to initialization.

## Supplementary Note 2. EXPERIMENTAL SETUP AND DATA GENERATION

Bare MSS chips were coated with 12 distinct polymer films. Polymer materials were dissolved in DMF or TCE to prepare solutions, which were deposited onto silicon membranes by inkjet spotting. The solvent was evaporated by heating on a hot plate at 60°C, followed by drying at room temperature for 24 hours. The coated chips were then placed in a sensor cell for gas exposure.

TABLE S1. Network architecture and training hyperparameters.

| **Parameter** | **Value** |
|---|---|
| *Architecture* | |
| TCN layers / channels / kernel size | 11 / 64 / 3 |
| Dilation pattern | $2^0, 2^1, \ldots, 2^{10}$ |
| Receptive field | 4095 time steps |
| Shared MLP hidden layers / dim | 2 / 512 |
| Prediction head hidden layers / dim | 1 / 128 |
| Dropout | 0.1 |
| *Training* | |
| Input length / sampling rate | 3200 pts (160 s) / 20 Hz |
| Normalization / batch size | z-score / 8 |
| Optimizer | Adan |
| Learning rate (initial → min) | $10^{-3} \rightarrow 10^{-5}$ |
| LR scheduler (factor / patience) | ReduceLROnPlateau (0.5 / 200) |
| Max epochs / early stopping patience | 2000 / 800 |
| $\omega_{C_g}$ / $\omega_V$ / MDMM damping | 10 / 1 / 10 |

Dynamic gas mixtures were generated using an in-house gas delivery system (Fig. S3) comprising three mass flow controllers (MFCs) with $N_2$ as the carrier gas. MFC-1 and MFC-2 were connected to bubblers containing liquid ethanol and water, respectively; the carrier gas evaporated the liquid samples into the gas phase. MFC-3 provided dilution $N_2$ to achieve the desired concentrations. The mixture composition was controlled by dynamically regulating the MFC flow rates. Sensor signals were acquired at 100 Hz and flow rates at 20 Hz using a LabVIEW-based data acquisition system.

Four dynamic waveform patterns were used to drive the MFCs: rectangular, sawtooth, reverse sawtooth, and Gaussian. These patterns approximate sensor exposure profiles from a steady source, an approaching source, a receding source, and a point source, respectively (Fig. S4).

Data were collected under two regimes: synchronous mixing, where both MFCs follow the same waveform to emulate a single source, and asynchronous mixing, where each MFC follows an independently selected waveform to emulate multiple sources (Fig. S5).

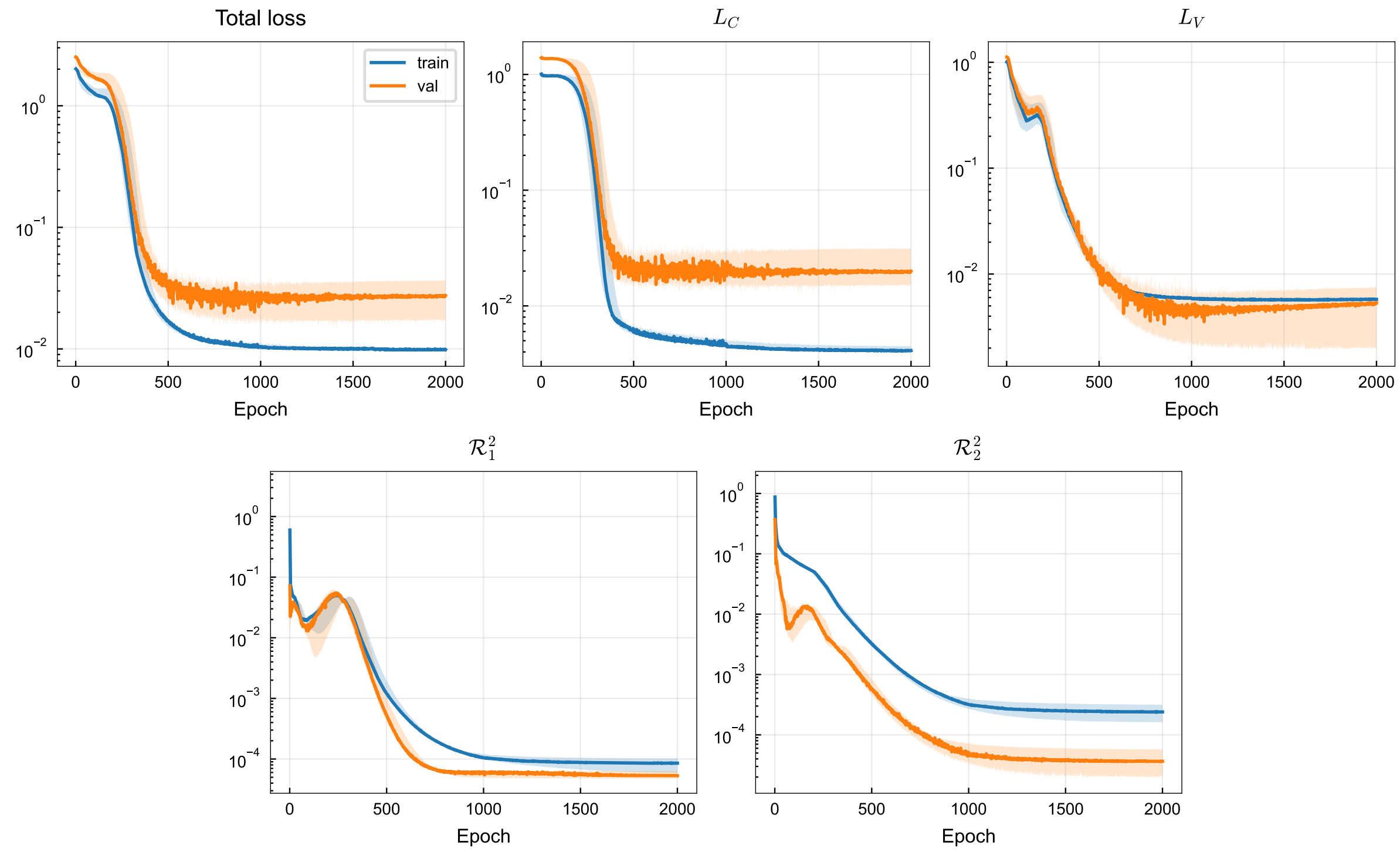


FIG. S2. Training and validation loss curves for the Physics model across five random seeds. Solid lines: median; shaded bands: min–max range. Top row: total loss, concentration loss $L_C$, and signal reconstruction loss $L_V$. Bottom row: squared physics residuals $\mathscr{R}_1^2$ (sorption) and $\mathscr{R}_2^2$ (viscoelastic).

Each measurement cycle lasts 240 s and comprises three phases: an initial purge (20 s), an active injection period (120 s), and a recovery purge (100 s). During the active injection period, 4–5 waveform segments of 10–20 s duration are interspersed with purge intervals of 10–20 s. Representative sensor output signals are shown in Fig. S6.

## 2.1. Training data generation

Synchronous training data were generated by systematically varying the waveform type and mixture composition. Pure-component sequences (single active MFC, rectangular waveform, concentration 0–30%) were generated for each gas species (5 sequences per species). Mixed-component sequences were generated by combining all four waveform types with three staggered concentration ranges for MFC-1 and MFC-2—e.g., (2–15%, 15–30%), (5–20%, 2–15%), and (10–25%, 5–20%)—with 5 repetitions per combination, yielding $4\times3\times5 = 60$ mixed sequences. In synchronous mode,

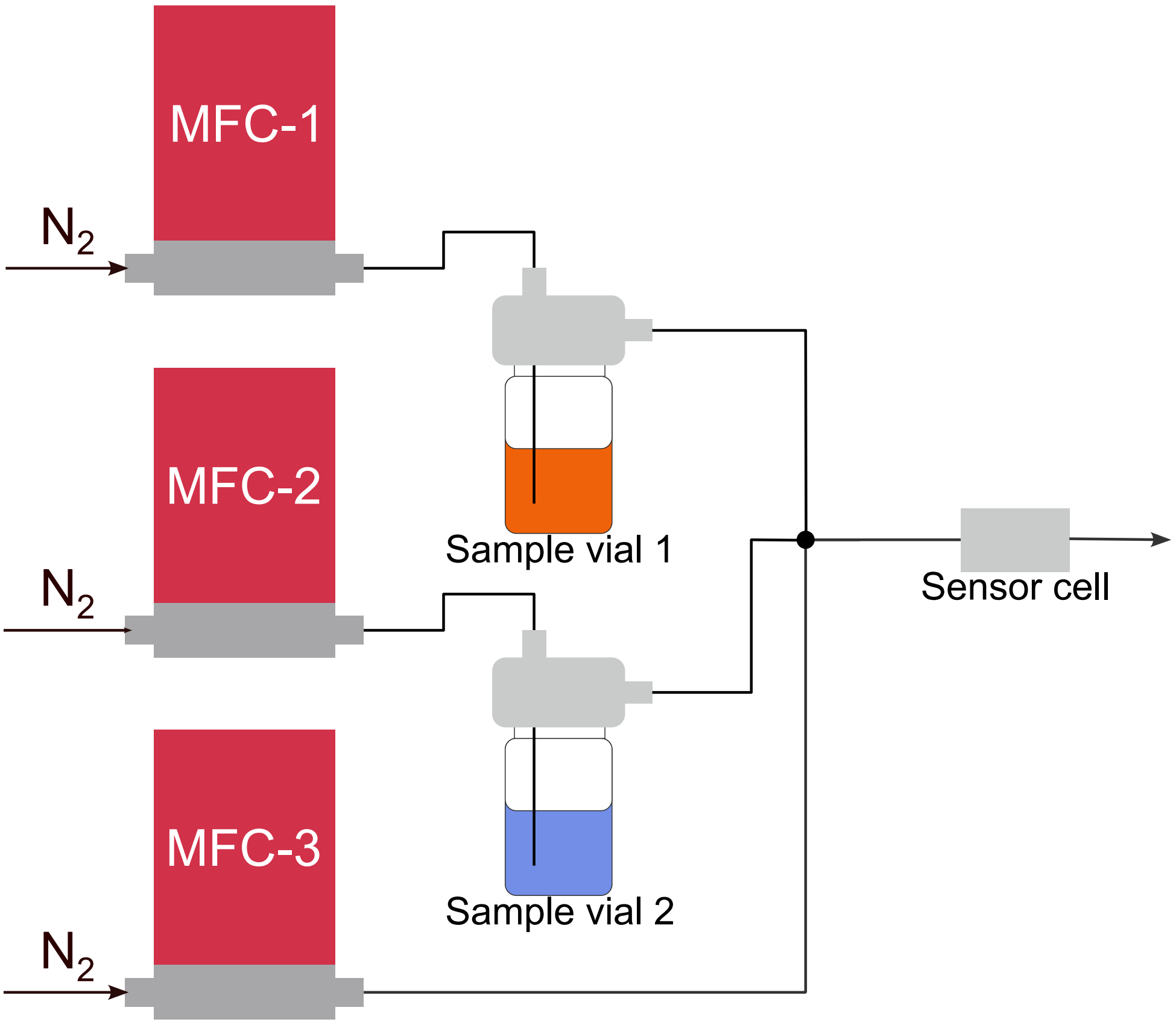


FIG. S3. Schematic of the experimental setup. MFC-1 and MFC-2 carry $N_2$ through bubblers containing ethanol and water, respectively. MFC-3 provides dilution $N_2$. The mixed gas is delivered to a sensor cell housing the 12-channel MSS array.

both MFCs share the same waveform type within each sequence. The synchronous training set thus comprises 70 sequences in total (10 pure-component + 60 mixed-component).

Asynchronous training data were generated analogously, except that each MFC was driven by an independently selected waveform type within each sequence, producing more complex multi-source mixing scenarios. This yields another 70 sequences.

### 2.2. Test data generation

Test sequences were generated with randomly selected waveform types, concentration ranges (0–30% for each component), and waveform durations. Each test sequence contains 3–5 waveform segments with at least three distinct waveform types. Synchronous and asynchronous test sets each comprise 30 sequences (with 2 warm-up sequences excluded).

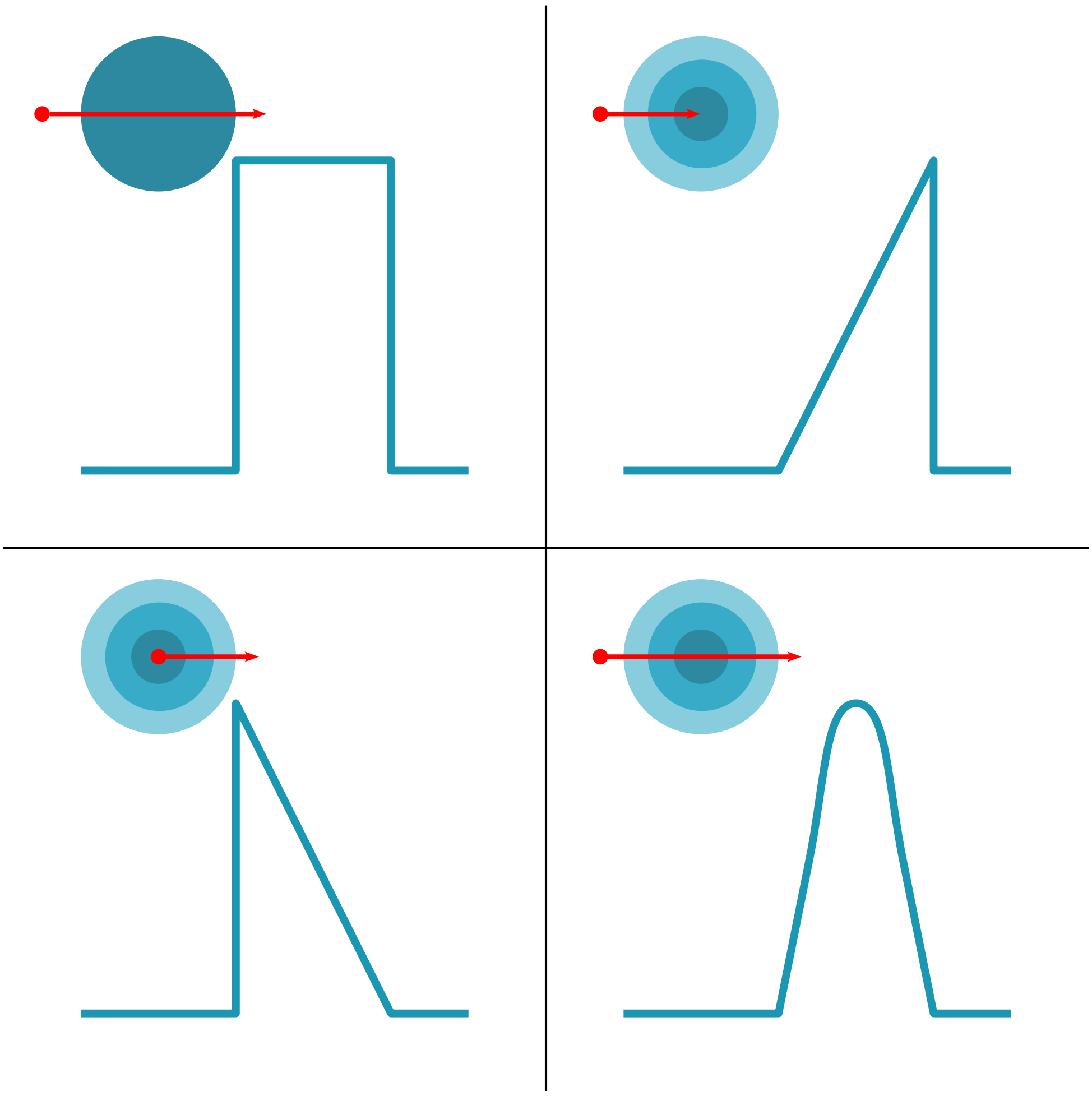

FIG. S4. The four waveform patterns used to drive the MFCs: rectangular, sawtooth, reverse sawtooth, and Gaussian.

### 2.3. Dataset split

The final dataset split is summarized in Table S2. To increase the diversity of training data, 10 synchronous test sequences were transferred to the training set. Each sequence is truncated to 160 s (3200 time steps at 20 Hz) for training, discarding the final recovery phase. All input signals are z-score normalized using training-set statistics; the same normalization parameters are applied to the test set without refitting.

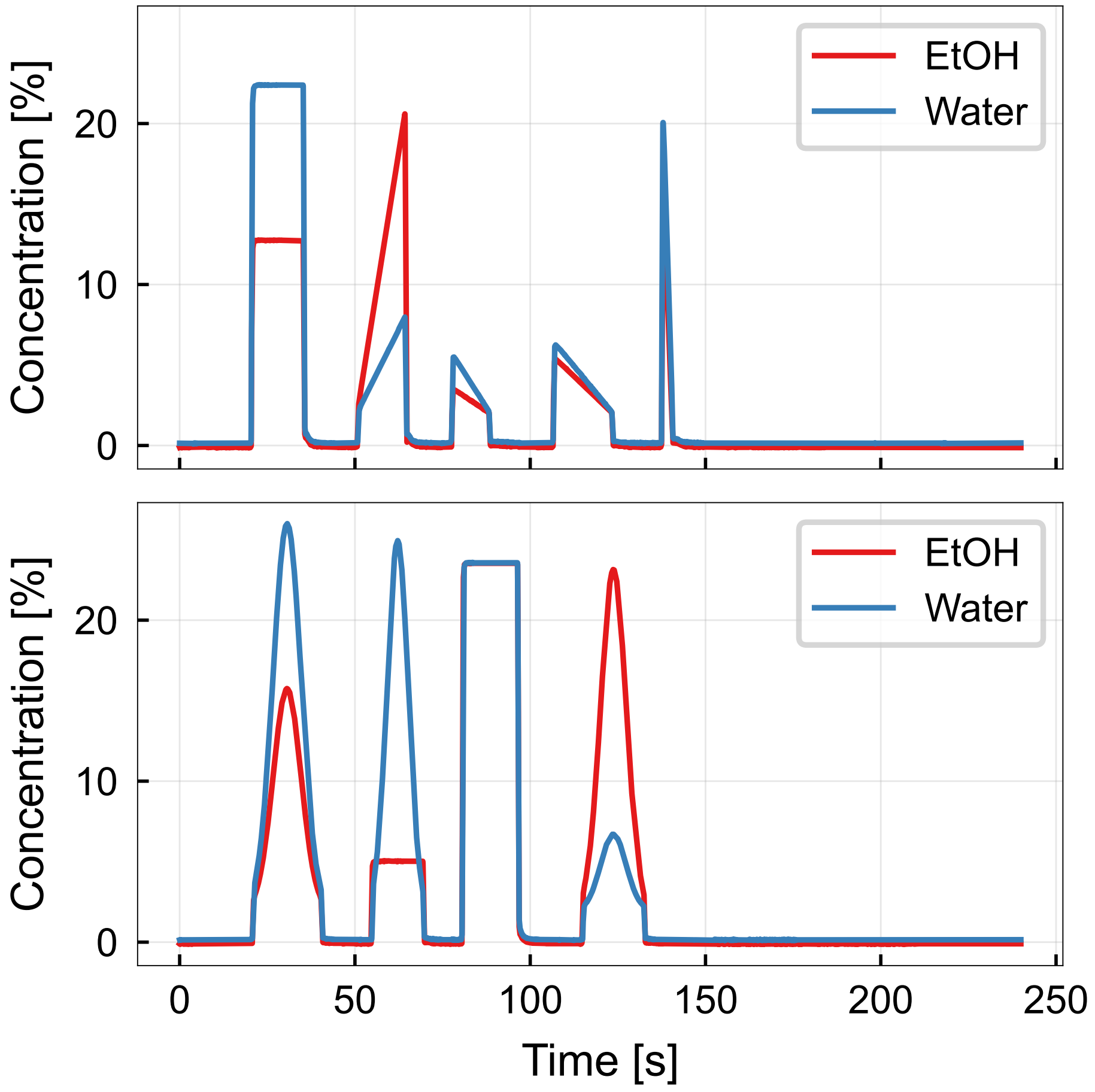


FIG. S5. Representative gas-phase concentration profiles $C_g$ under synchronous (upper) and asynchronous (lower) mixing. In the synchronous regime, both components share the same waveform; in the asynchronous regime, each component follows an independent waveform.

## Supplementary Note 3. PHYSICAL MODEL AND NORMALIZATION

This section presents the theoretical foundations of the gas–solid interaction dynamics in the membrane-type surface stress sensor (MSS). The forward problem for a single sensor has been studied for the single-component case by Imamura *et al.*[4] and Minami *et al.*[5], and for the multi-component case by Feng *et al.*[6] and Minami *et al.*[7]. The forward process comprises two stages: sorption kinetics and viscoelastic coupling.

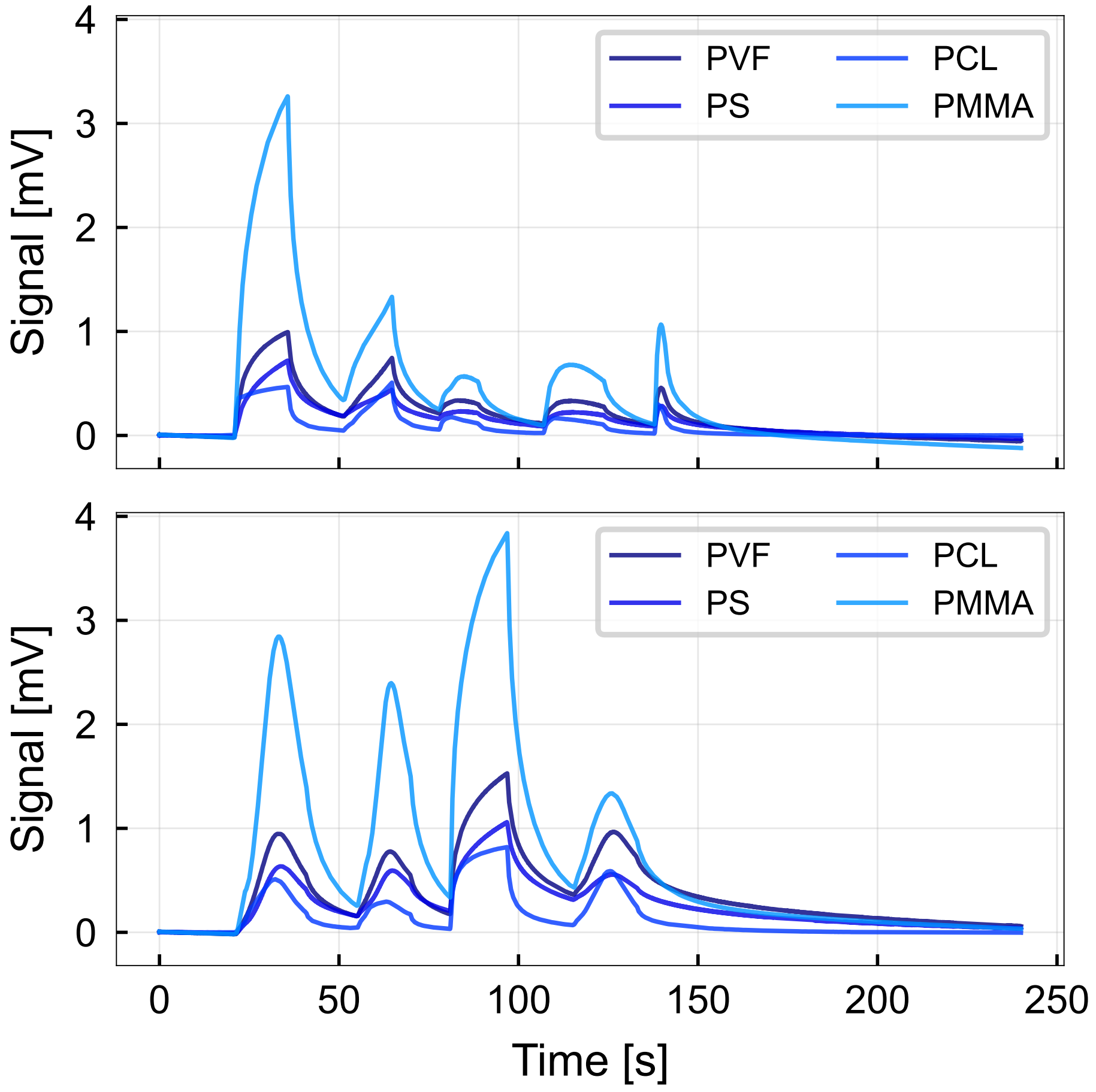


FIG. S6. Representative 12-channel sensor output signals under synchronous (upper) and asynchronous (lower) mixing, corresponding to the concentration profiles in Fig. S5.

TABLE S2. Dataset split. Numbers in parentheses indicate the original allocation before transfer.

| | Training | Test |
|---|---|---|
| Synchronous | 80 (70 + 10 transferred) | 20 (30 − 10 transferred) |
| Asynchronous | 70 | 30 |
| Total | 150 | 50 |

### 3.1. Sorption kinetics

Mass transport from the gas phase to the solid phase is governed by Henry's law and Fick's law. Under the assumption of pseudo-first-order sorption kinetics, the film-phase concentration evolves

as

$$\frac{dC_f}{dt} = \frac{1}{\tau_s}(K_p C_g - C_f) \tag{S6}$$

where $\tau_s$ is the sorption time constant, $K_p$ is the partition coefficient, $C_f$ is the film-phase concentration, and $C_g$ is the gas-phase concentration. The parameter $\tau_s$ governs the transient response, while $K_p$ determines the equilibrium sorption capacity. Both parameters depend on the physical properties of the receptor material and the analyte, and are treated as constants for a given sensor–gas pair.

### 3.2. Viscoelastic coupling

The sensing layer swells upon gas absorption. Under the small-strain assumption for isotropic swelling, the linear strain $\varepsilon$ is

$$\varepsilon = \frac{v}{3} C_f \tag{S7}$$

where $v$ is the specific volume of the absorbed gas. For a viscoelastic material described by the standard linear solid model, the stress–strain relationship is

$$\tau_r E_U \frac{d\varepsilon}{dt} + E_R \varepsilon = \tau_r \frac{d\sigma}{dt} + \sigma \tag{S8}$$

where $E_U$ is the unrelaxed modulus, $E_R$ is the relaxed modulus, and $\tau_r$ is the mechanical relaxation time constant. The sensor output signal is

$$V = \gamma \sigma \tag{S9}$$

where $\gamma$ is the stress-to-electrical conversion factor. In this work $\gamma$ is absorbed into the normalization (effectively $\gamma = 1$), so that $V \equiv \sigma$; the main text uses $\sigma$ throughout.

### 3.3. Multi-gas, multi-sensor extension

The single-sensor formulation generalizes to an array of $n_{\text{sensors}}$ sensors exposed to a mixture of $n_{\text{gases}}$ gases. Let $C_{g,i}$ denote the gas-phase concentration of the $i$-th component and $C_{f,ij}$ the corresponding film-phase concentration in the $j$-th sensor.

Assuming negligible inter-component interactions in the film phase and no sorption-induced aging, the sorption kinetics for each gas–sensor pair are

$$\frac{dC_{f,ij}}{dt} = \frac{1}{\tau_{s,ij}}(K_{p,ij} C_{g,i} - C_{f,ij}) \tag{S10}$$

where $\tau_{s,ij}$ and $K_{p,ij}$ are the sorption time constant and partition coefficient for the $i$-th gas in the $j$-th sensor. In matrix form,

$$\dot{\mathbf{C}}_f = \mathbf{T}_s^{-1} \odot (\mathbf{K}_p \mathbf{C}_g - \mathbf{C}_f) \tag{S11}$$

where $\mathbf{C}_g \in \mathbb{R}^{n_{\text{gases}}}$, $\mathbf{C}_f \in \mathbb{R}^{n_{\text{sensors}} \times n_{\text{gases}}}$, $\mathbf{T}_s \in \mathbb{R}^{n_{\text{sensors}} \times n_{\text{gases}}}$, and $\mathbf{K}_p \in \mathbb{R}^{n_{\text{sensors}} \times n_{\text{gases}}}$.

The strain in the $j$-th sensor is

$$\varepsilon_j = \frac{v}{3} \sum_{i=1}^{n_{\text{gases}}} C_{f,ij} \tag{S12}$$

or equivalently $\boldsymbol{\varepsilon} = (v/3)\, \mathbf{1}^{\top} \mathbf{C}_f$, where $\mathbf{1} \in \mathbb{R}^{n_{\text{gases}}}$ is the all-ones vector and $\boldsymbol{\varepsilon} \in \mathbb{R}^{n_{\text{sensors}}}$.

The stress–strain relationship for each sensor follows the standard linear solid model:

$$\tau_{r,j} E_{U,j} \frac{d\varepsilon_j}{dt} + E_{R,j} \varepsilon_j = \tau_{r,j} \frac{d\sigma_j}{dt} + \sigma_j \tag{S13}$$

In matrix form,

$$(\boldsymbol{\tau}_r \odot \mathbf{E}_U) \odot \dot{\boldsymbol{\varepsilon}} + \mathbf{E}_R \odot \boldsymbol{\varepsilon} = \boldsymbol{\tau}_r \odot \dot{\boldsymbol{\sigma}} + \boldsymbol{\sigma} \tag{S14}$$

where $\boldsymbol{\tau}_r, \mathbf{E}_U, \mathbf{E}_R, \boldsymbol{\varepsilon}, \boldsymbol{\sigma} \in \mathbb{R}^{n_{\text{sensors}}}$.

The output signal of the $j$-th sensor is $V_j = \gamma \sigma_j$, or in vector form $\mathbf{V} = \gamma \boldsymbol{\sigma}$ with $\mathbf{V} \in \mathbb{R}^{n_{\text{sensors}}}$. As noted above, $\gamma = 1$ is assumed throughout, so $\mathbf{V} \equiv \boldsymbol{\sigma}$.

### 3.4. Normalization and parameter recovery

For numerical stability and faster convergence during neural network training, all variables are standardized[8]. The normalized variable $X'$ is related to the original variable $X$ by

$$X' = \frac{X - \mu_X}{\sigma_X} \tag{S15}$$

where $\mu_X$ and $\sigma_X$ are the mean and standard deviation of $X$. The normalized variables include $C'_{g,i}$, $C'_{f,ij}$, $\varepsilon'_j$, $\sigma'_j$, and $V'_j$:

$$\begin{cases} C'_{g,i} = \dfrac{C_{g,i} - \mu_{g,i}}{\sigma_{g,i}} & \text{(S16a)} \\[2ex] C'_{f,ij} = \dfrac{C_{f,ij} - \mu_{f,ij}}{\sigma_{f,ij}} & \text{(S16b)} \end{cases}$$

$$\begin{cases} \varepsilon'_j = \dfrac{\varepsilon_j - \mu_{\varepsilon,j}}{\sigma_{\varepsilon,j}} & \text{(S17a)} \\[2ex] \sigma'_j = \dfrac{\sigma_j - \mu_{\sigma,j}}{\sigma_{\sigma,j}} & \text{(S17b)} \end{cases}$$

Substituting Eqs. S16 and S17 into the governing equations S10 and S13 yields the normalized sorption kinetics:

$$\frac{dC'_{f,ij}}{dt} = A_{ij}C'_{g,i} - B_{ij}C'_{f,ij} + C_{ij} \tag{S18}$$

with

$$\begin{cases} A_{ij} = \dfrac{K_{p,ij}\sigma_{g,i}}{\tau_{s,ij}\sigma_{f,ij}} & \text{(S19a)} \\ B_{ij} = \dfrac{1}{\tau_{s,ij}} & \text{(S19b)} \\ C_{ij} = \dfrac{K_{p,ij}\mu_{g,i} - \mu_{f,ij}}{\tau_{s,ij}\sigma_{f,ij}} & \text{(S19c)} \end{cases}$$

where $A_{ij}$, $B_{ij}$, $C_{ij}$ are learnable parameters that implicitly encode the sorption rate constant $\tau_{s,ij}^{-1}$, partition coefficient $K_{p,ij}$, and the normalization statistics $(\mu_{g,i}, \mu_{f,ij}, \sigma_{g,i}, \sigma_{f,ij})$.

The normalized viscoelastic equation for each sensor $j$ is

$$D_j\frac{d\varepsilon'_j}{dt} + E_j\varepsilon'_j = F_j\frac{d\sigma'_j}{dt} + G_j\sigma'_j + H_j \tag{S20}$$

with

$$\begin{cases} D_j = \tau_{r,j}E_{U,j}\sigma_{\varepsilon,j} & \text{(S21a)} \\ E_j = E_{R,j}\sigma_{\varepsilon,j} & \text{(S21b)} \\ F_j = \tau_{r,j}\sigma_{\sigma,j} & \text{(S21c)} \\ G_j = \sigma_{\sigma,j} & \text{(S21d)} \\ H_j = \mu_{\sigma,j} - E_{R,j}\mu_{\varepsilon,j} & \text{(S21e)} \end{cases}$$

where $D_j$, $E_j$, $F_j$, $G_j$, $H_j$ are learnable parameters encoding the relaxation time constant $\tau_{r,j}$, unrelaxed modulus $E_{U,j}$, relaxed modulus $E_{R,j}$, and the normalization statistics $(\mu_{\varepsilon,j}, \mu_{\sigma,j}, \sigma_{\varepsilon,j}, \sigma_{\sigma,j})$. The following physical parameters can be recovered from the composite parameters:

$$\begin{cases} \tau_{r,j} = \dfrac{F_j}{G_j} & \text{(S22a)} \\ \dfrac{E_{U,j}}{E_{R,j}} = \dfrac{D_jG_j}{E_jF_j} & \text{(S22b)} \end{cases}$$

### 3.5. Recovered physical parameters

Figure S7 shows the physical parameters recovered from the composite learnable coefficients (Eq. S22), aggregated over five independent training runs with different random seeds. The sorption time constant $\tau_s$ exhibits channel-dependent values on the order of 1–5 s, with the two gas

species (ethanol and water) showing distinct time scales in most channels, consistent with the expected difference in diffusion kinetics. Most channels show small cross-seed variance, indicating robust parameter recovery. However, two channels—PS and PVC—exhibit markedly larger variance (standard deviation ∼3–4 s). This instability may reflect either limited parameter identifiability, whereby the optimization landscape admits multiple $\tau_s$ values with similar loss, or physical drift in the receptor film properties during extended measurement campaigns, which would introduce sequence-to-sequence variability that manifests as seed-dependent optima.

The modulus ratio $E_U/E_R$ converges to approximately unity across all channels and all seeds, with negligible variance ($< 0.2\%$). This result indicates that the standard linear solid (SLS) model effectively reduces to a purely elastic response under the present experimental conditions. The physical origin of this behavior lies in the cascade of sorption kinetics and viscoelastic coupling: the gas-phase concentration $C_g(t)$ is first filtered by first-order sorption dynamics with time constant $\tau_s$, which attenuates fast temporal variations before they reach the mechanical element. Among the four stimulus waveforms employed (rectangular, sawtooth, reverse sawtooth, and Gaussian), only the rectangular waveform produces near-instantaneous changes in $C_g$; the remaining waveforms impose inherently slower concentration ramps. After sorption filtering, the resulting strain rate $\dot{\varepsilon}$ is sufficiently low that the SLS response becomes indistinguishable from the elastic limit ($E_U = E_R$). Previous studies on sorption-coupled SLS systems[4,5] employed exclusively rectangular gas stimuli, where the fast loading rate produced clearly resolvable viscoelastic effects. In the present work, the rectangular waveform constitutes only a fraction of the training data, and the viscoelastic signature it produces is insufficient to constrain $E_U/E_R$ away from the elastic limit during optimization.

The mechanical relaxation time constant $\tau_r$ shows moderate channel-to-channel variation (0.1–0.6 s) with relatively small cross-seed variance, indicating that this parameter is reasonably well constrained despite the limited viscoelastic contrast. The numerical values are listed in Table S3.

## Supplementary Note 4. ABLATION STUDIES ON PHYSICAL CONSTRAINTS

To evaluate the contribution of physical constraints, three model variants are compared: (i) **Physics**—the full model with both sorption-kinetic and viscoelastic constraints enforced via MDMM; (ii) **Sorp-only**—retaining only the sorption-kinetic constraint (Eq. S1) and removing the viscoelastic constraint (Eq. S2); and (iii) **Data-only**—removing all physical constraints and training with the data loss alone. All three variants share the same network architecture and training

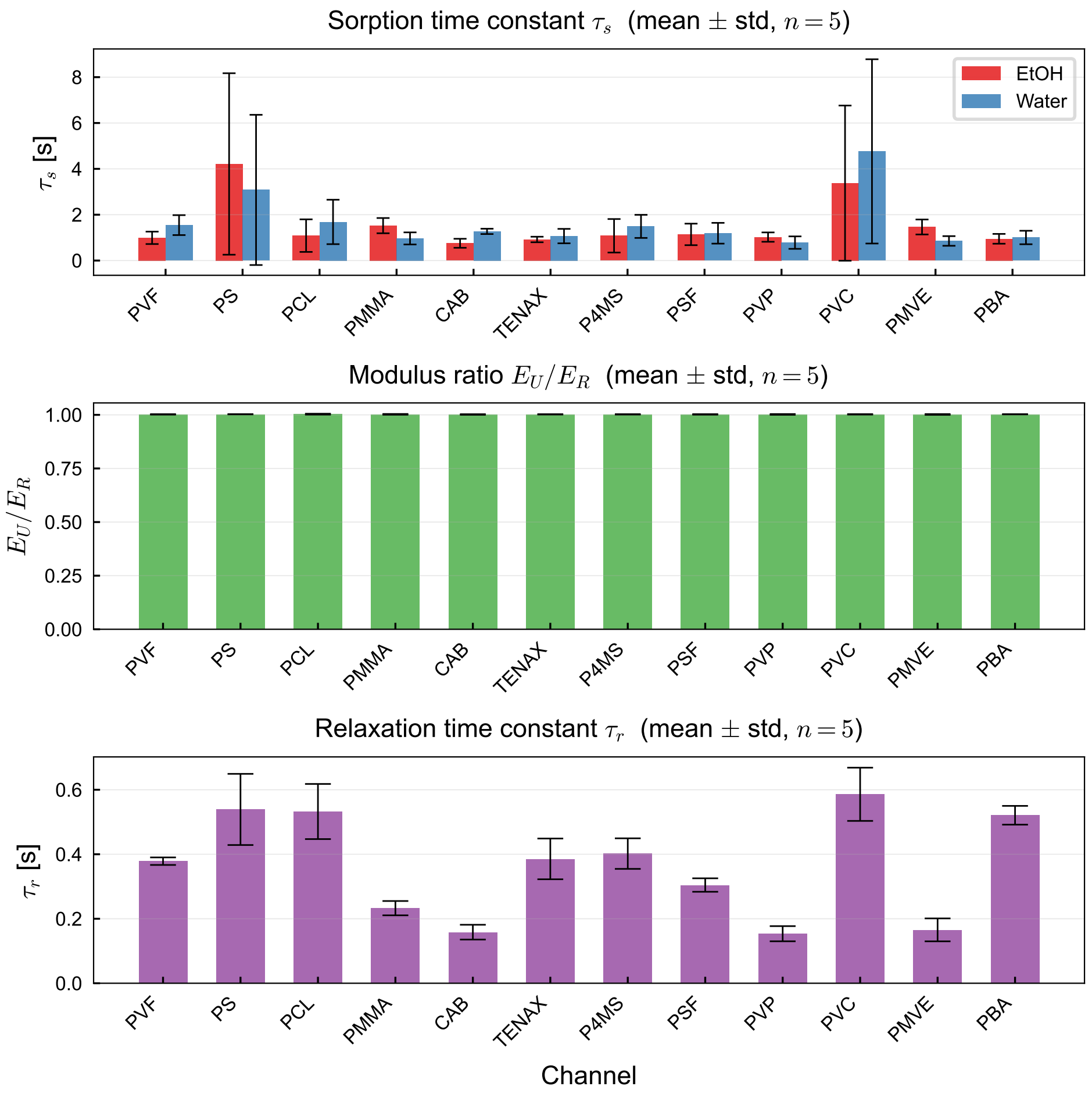


FIG. S7. Recovered physical parameters for each sensor channel, aggregated over five independent training runs (mean ± std). Top: sorption time constant $\tau_s$ for ethanol and water. Middle: modulus ratio $E_U/E_R$. Bottom: mechanical relaxation time constant $\tau_r$.

hyperparameters; only the constraint configuration differs.

### 4.1. Normal-input reconstruction

Under normal test inputs (Fig. S8), all three models reconstruct gas-phase concentrations $\hat{C}_g$ with comparable accuracy. The reconstructed signal $\hat{\sigma}$ also closely tracks the input signal across models. The primary difference appears in the film-phase concentration $\hat{C}_f$: the Physics model produces smooth, physically consistent $\Delta C_f$ trajectories that follow the expected sorption–desorption dynamics, while the Data-only model produces $\Delta C_f$ trajectories that lack the characteristic time

TABLE S3. Recovered physical parameters (mean ± std over 5 training runs). The modulus ratio $E_U/E_R$ converges to $1.002 \pm 0.001$ for all channels (see text) and is omitted from the table.

| Channel | $\tau_{s,\mathrm{EtOH}}$ (s) | $\tau_{s,\mathrm{Water}}$ (s) | $\tau_r$ (s) |
|---|---|---|---|
| PVF | $0.99 \pm 0.27$ | $1.55 \pm 0.43$ | $0.38 \pm 0.01$ |
| PS | $4.21 \pm 3.96$ | $3.08 \pm 3.28$ | $0.54 \pm 0.11$ |
| PCL | $1.09 \pm 0.71$ | $1.68 \pm 0.97$ | $0.53 \pm 0.09$ |
| PMMA | $1.52 \pm 0.33$ | $0.97 \pm 0.26$ | $0.23 \pm 0.02$ |
| CAB | $0.75 \pm 0.20$ | $1.27 \pm 0.12$ | $0.16 \pm 0.02$ |
| TENAX | $0.92 \pm 0.12$ | $1.07 \pm 0.31$ | $0.39 \pm 0.06$ |
| P4MS | $1.08 \pm 0.73$ | $1.49 \pm 0.50$ | $0.40 \pm 0.05$ |
| PSF | $1.14 \pm 0.47$ | $1.19 \pm 0.45$ | $0.30 \pm 0.02$ |
| PVP | $1.02 \pm 0.20$ | $0.78 \pm 0.27$ | $0.15 \pm 0.02$ |
| PVC | $3.38 \pm 3.38$ | $4.76 \pm 4.02$ | $0.59 \pm 0.08$ |
| PMVE | $1.46 \pm 0.33$ | $0.85 \pm 0.21$ | $0.17 \pm 0.04$ |
| PBA | $0.95 \pm 0.21$ | $1.01 \pm 0.29$ | $0.52 \pm 0.03$ |

constants of diffusion-limited sorption.

## 4.2. Baseline distributions of self-verification metrics

The distributions of the signal inconsistency $I_\sigma$, sorption residual $\overline{|\mathscr{R}_1|}$, and viscoelastic residual $\overline{|\mathscr{R}_2|}$ computed on 50 normal test sequences reveal systematic differences among the three models (Fig. S9). All three models achieve similar $I_\sigma$ values ($10^{-3}$–$10^{-2}$), confirming that signal reconstruction accuracy is largely determined by the data loss rather than the physical constraints. In contrast, $\overline{|\mathscr{R}_1|}$ and $\overline{|\mathscr{R}_2|}$ differ by orders of magnitude: the Physics model achieves the lowest residuals, the Sorp-only model shows intermediate $\overline{|\mathscr{R}_1|}$ but elevated $\overline{|\mathscr{R}_2|}$ (reflecting the absence of the viscoelastic constraint), and the Data-only model exhibits $\overline{|\mathscr{R}_1|}$ approximately three orders of magnitude higher. These distributions serve as baselines for interpreting the perturbation experiments below.

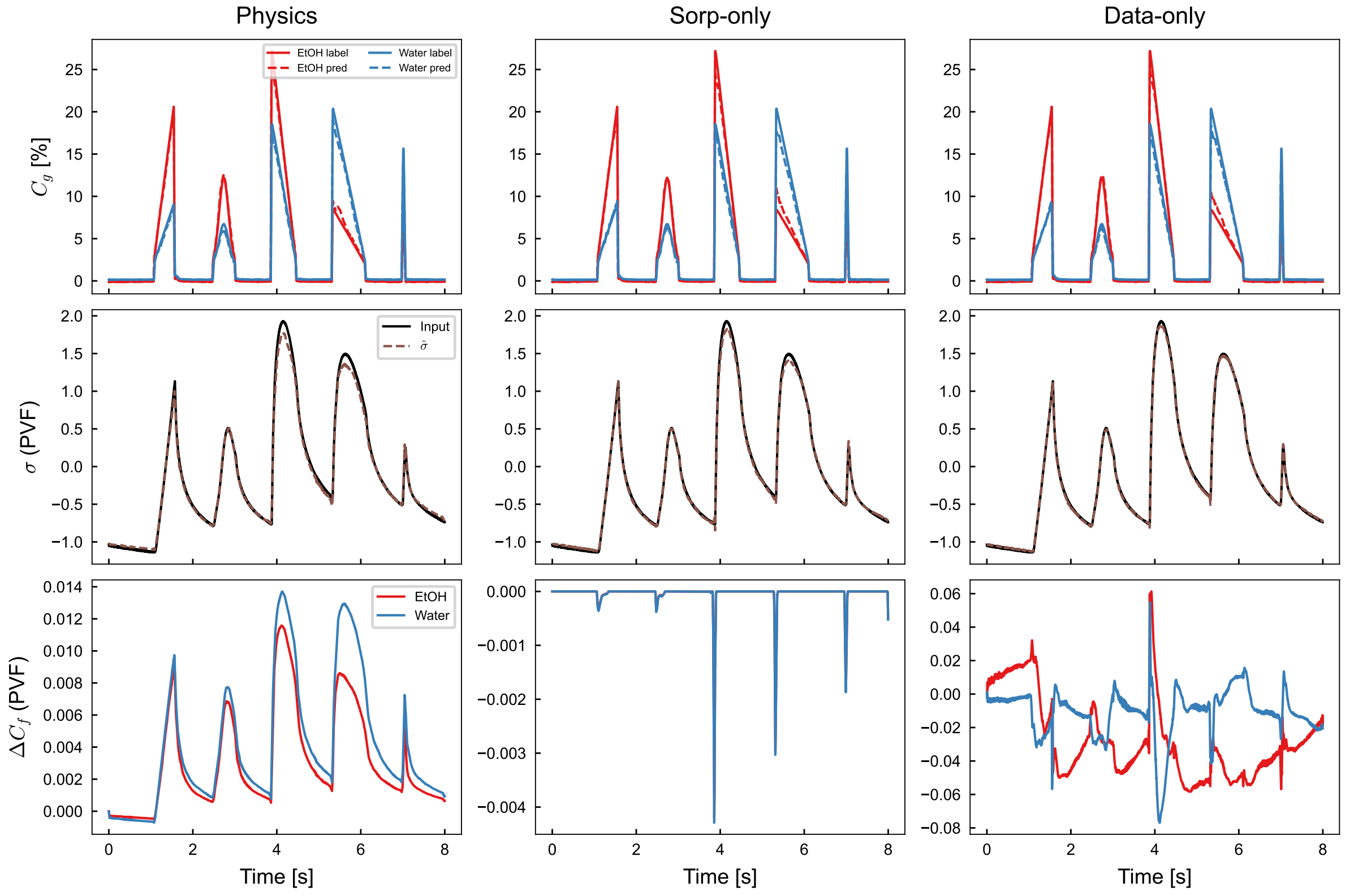


FIG. S8. Reconstruction of a representative test sequence by the three model variants. Top row: $C_g$ predictions (dashed) overlaid on ground truth (solid) for ethanol and water. Middle row: reconstructed signal $\hat{\sigma}$ versus input for one representative channel (PVF). Bottom row: change in film-phase concentration $\Delta C_f = C_f(t) - C_f(0)$ for both gas species.

### 4.3. Signal inconsistency under input perturbations

To test the sensitivity of $I_\sigma$ as a self-verification metric, perturbations are applied to individual sensor channels in the normalized signal space. Two perturbation types are examined: additive offset ($\sigma'_j \to \sigma'_j + \delta$, $\delta \in [0, 1]$) and multiplicative gain reduction ($\sigma'_j \to \alpha\sigma'_j$, $\alpha \in [0.2, 1]$).

For offset perturbations (Fig. S10), $I_\sigma$ increases monotonically with the offset magnitude across all 12 channels and all three models. The Physics model shows the steepest increase, consistent with its tighter coupling between the reconstructed signal $\hat{\sigma}$ and the physical constraints. The inset KDE distributions at $\delta = 0$ indicate the baseline level for each model.

For gain reduction (Fig. S11), $I_\sigma$ increases as the scale factor decreases from 1.0 (unperturbed). The trend is qualitatively similar across models: reducing the signal amplitude drives $\hat{\sigma}$ away from

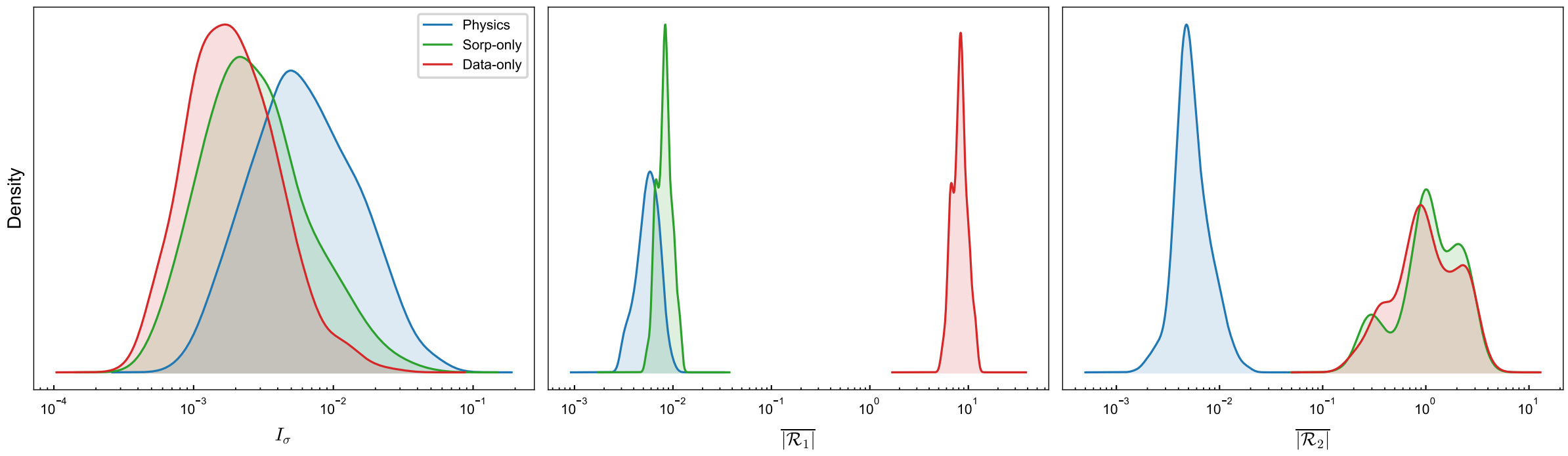


FIG. S9. Kernel density estimates of the self-verification metrics on 50 normal test sequences. Left: signal inconsistency $I_\sigma$. Center: mean sorption residual $\overline{|\mathscr{R}_1|}$. Right: mean viscoelastic residual $\overline{|\mathscr{R}_2|}$. Values are computed per channel per sequence (600 values per model).

the attenuated input, raising $I_\sigma$. The baseline KDE distributions at $\alpha = 1$ again confirm the unperturbed reference levels.

### 4.4. Physical residual stability under input perturbations

In contrast to the monotonic increase of $I_\sigma$, the physical residuals $\mathscr{R}_1$ and $\mathscr{R}_2$ remain comparatively stable across the same perturbation range, confirming that the model projects corrupted inputs onto the physical manifold rather than distorting its internal dynamics.

Figure S12 shows the sorption residual $\mathscr{R}_1$ as a function of offset and gain perturbation magnitude. Under offset perturbations (a, b), $\mathscr{R}_1$ exhibits only a mild increase; the perturbed and unperturbed channel distributions largely overlap. Under gain perturbations (c, d), $\mathscr{R}_1$ remains essentially flat. The viscoelastic residual $\mathscr{R}_2$ (Fig. S13) shows a similar pattern: a modest increase under offset perturbations and a slight decrease under gain perturbations. In both cases, the changes are orders of magnitude smaller than the corresponding $I_\sigma$ response (Figs. S10–S11).

### 4.5. Complementarity of $I_\sigma$ and physical residuals

The signal inconsistency $I_\sigma$ and the physical residuals $\mathscr{R}_1$, $\mathscr{R}_2$ capture different aspects of anomalous inputs. To demonstrate their complementarity, a same-sequence deletion splice is performed: 100 data points (5 s) are removed from the middle of a normal test sequence at $t = 85$ s, and the remaining segments are stitched together. The spliced signal is continuous and plausible

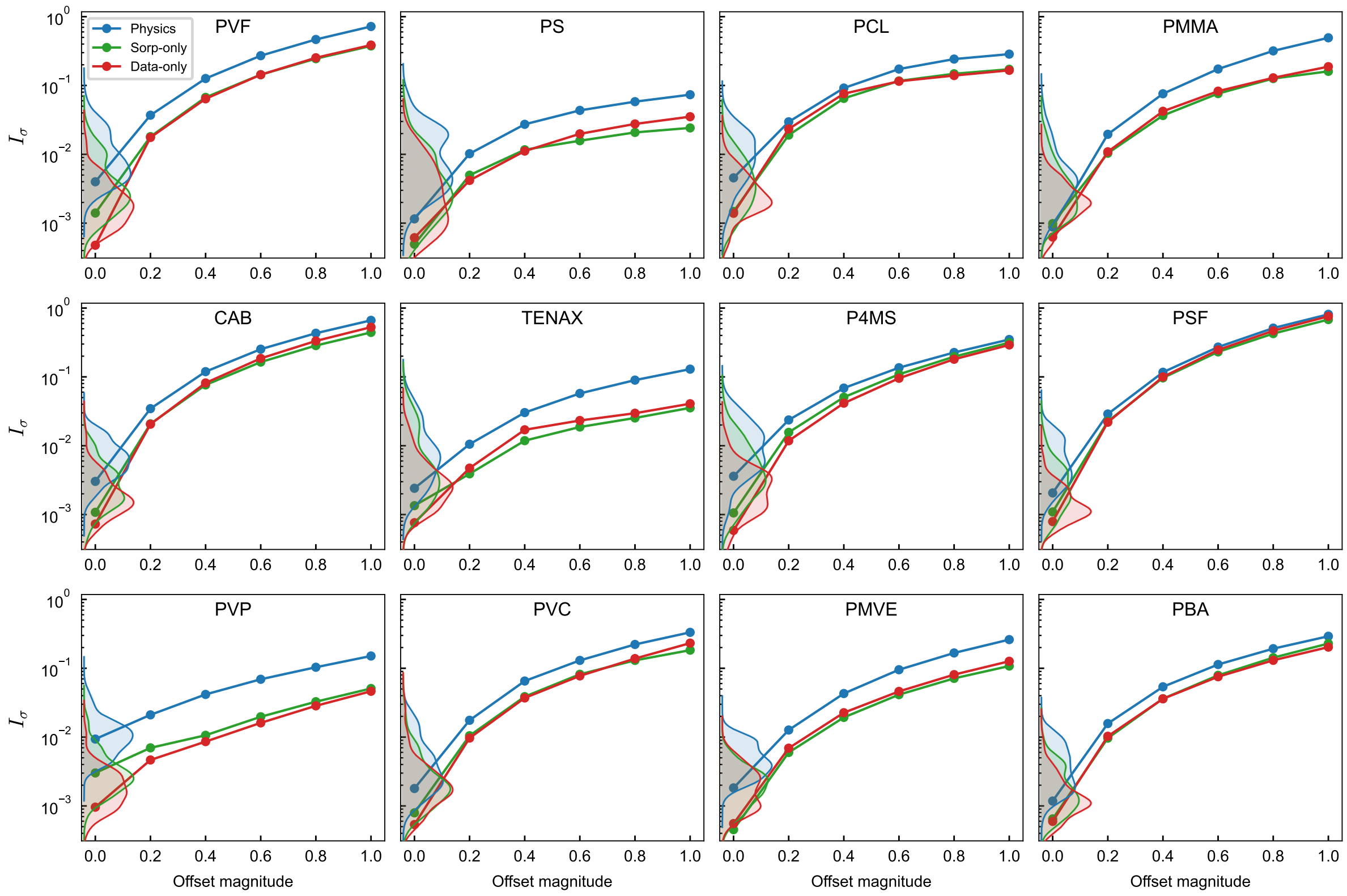


FIG. S10. Signal inconsistency $I_\sigma$ as a function of additive offset magnitude for each of the 12 sensor channels. Inset KDE distributions at offset = 0 show the baseline $I_\sigma$ for each model.

in appearance, but violates the underlying temporal dynamics at the splice point.

Figure S14 compares the Physics model's response to the spliced and normal sequences. The spliced input produces no substantial increase in the time-resolved $I_\sigma(t)$, because the signal values on either side of the splice are individually within the training distribution. However, the physical residuals $\mathscr{R}_1$ and $\mathscr{R}_2$ exhibit sharp spikes at the splice point ($t \approx 85$ s), because the abrupt transition violates the governing differential equations. The inset KDE distributions computed from normal test data confirm that the spike magnitudes far exceed the baseline distribution, while $I_\sigma(t)$ remains within normal range.

This result shows that $I_\sigma$ functions as a detector of global distributional shifts (offsets, gain changes), whereas $\mathscr{R}_1$ and $\mathscr{R}_2$ detect local violations of physical consistency—even when the input appears statistically normal. Together, they form a two-layer self-verification mechanism: $I_\sigma$ flags whether the input lies within the observed signal manifold, and $\mathscr{R}_1/\mathscr{R}_2$ flag whether the reconstructed dynamics are internally consistent.

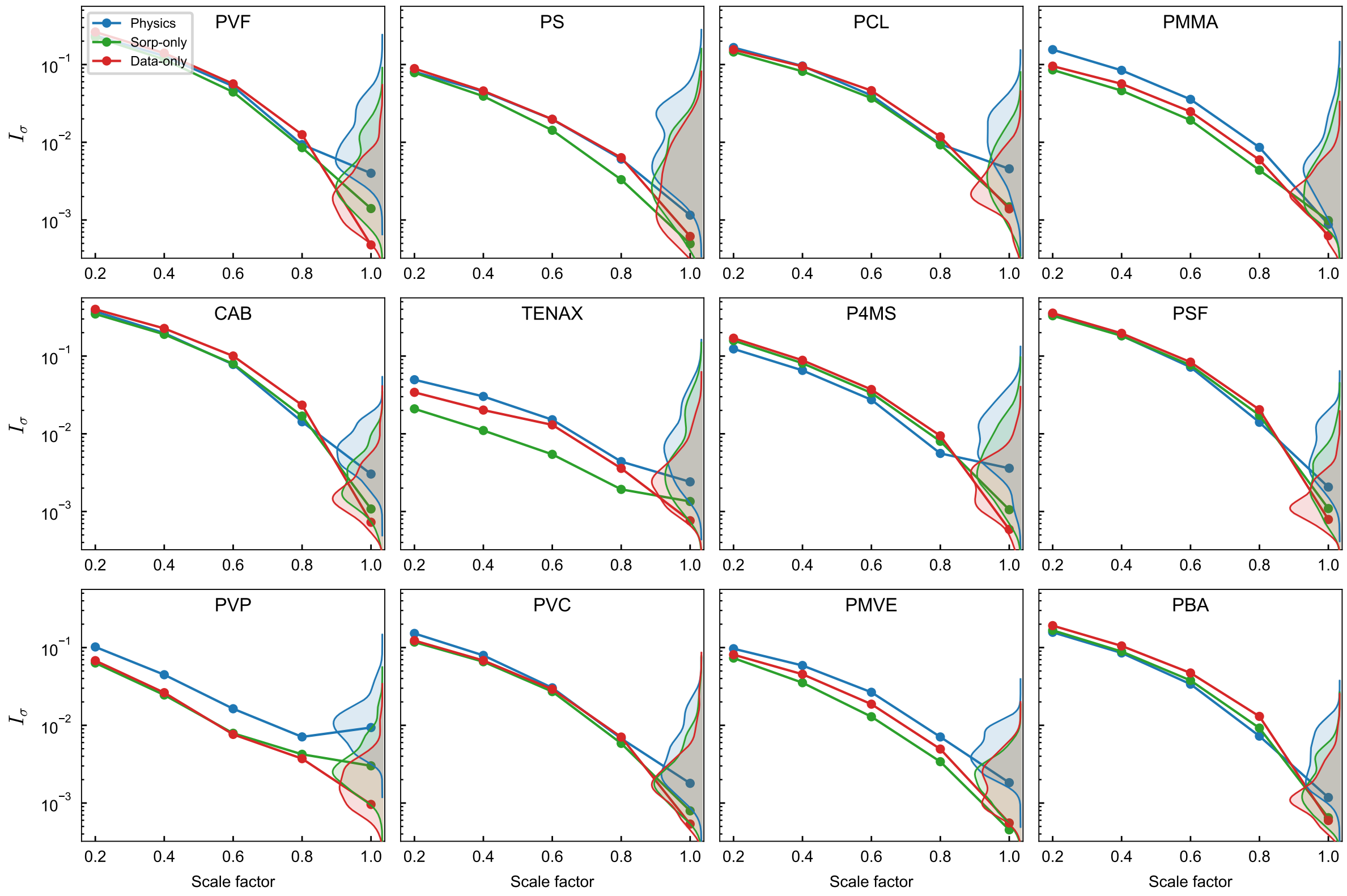


FIG. S11. Signal inconsistency $I_\sigma$ as a function of multiplicative scale factor for each of the 12 sensor channels. Inset KDE distributions at scale factor = 1 show the baseline $I_\sigma$ for each model.

A cross-sequence substitution variant extends this experiment to statistically camouflaged edits (Fig. 5e,f of the main text). The 70–90 s segment of each test sequence is replaced by the corresponding segment of the next sequence in the test set, so that every sample of the record is genuine measured data. Because all sequences share the same exposure phase structure, the substituted segment is statistically aligned with its surroundings. Figure S15 shows a representative example. The reconstructed $\hat{\sigma}$ tracks the substituted input without visible mismatch, and the test-set mean $I_\sigma$ increases by only 6.6% relative to the unperturbed baseline ($9.37 \times 10^{-3}$ versus $8.79 \times 10^{-3}$ in z-score normalized units), well within the baseline spread. The physical residuals $\mathscr{R}_1$ and $\mathscr{R}_2$, in contrast, exhibit sharp spikes at both junctions, bracketing the substituted region.

The main text reports residuals for a representative channel. Figure S16 summarizes junction detection across all twelve channels. For each channel and test sequence, the peak $|\mathscr{R}|$ within $\pm 1$ s of the junctions is compared with the 99th percentile of $|\mathscr{R}|$ over the rest of the record, a stringent baseline that includes the genuine residual transients accompanying sharp concentration

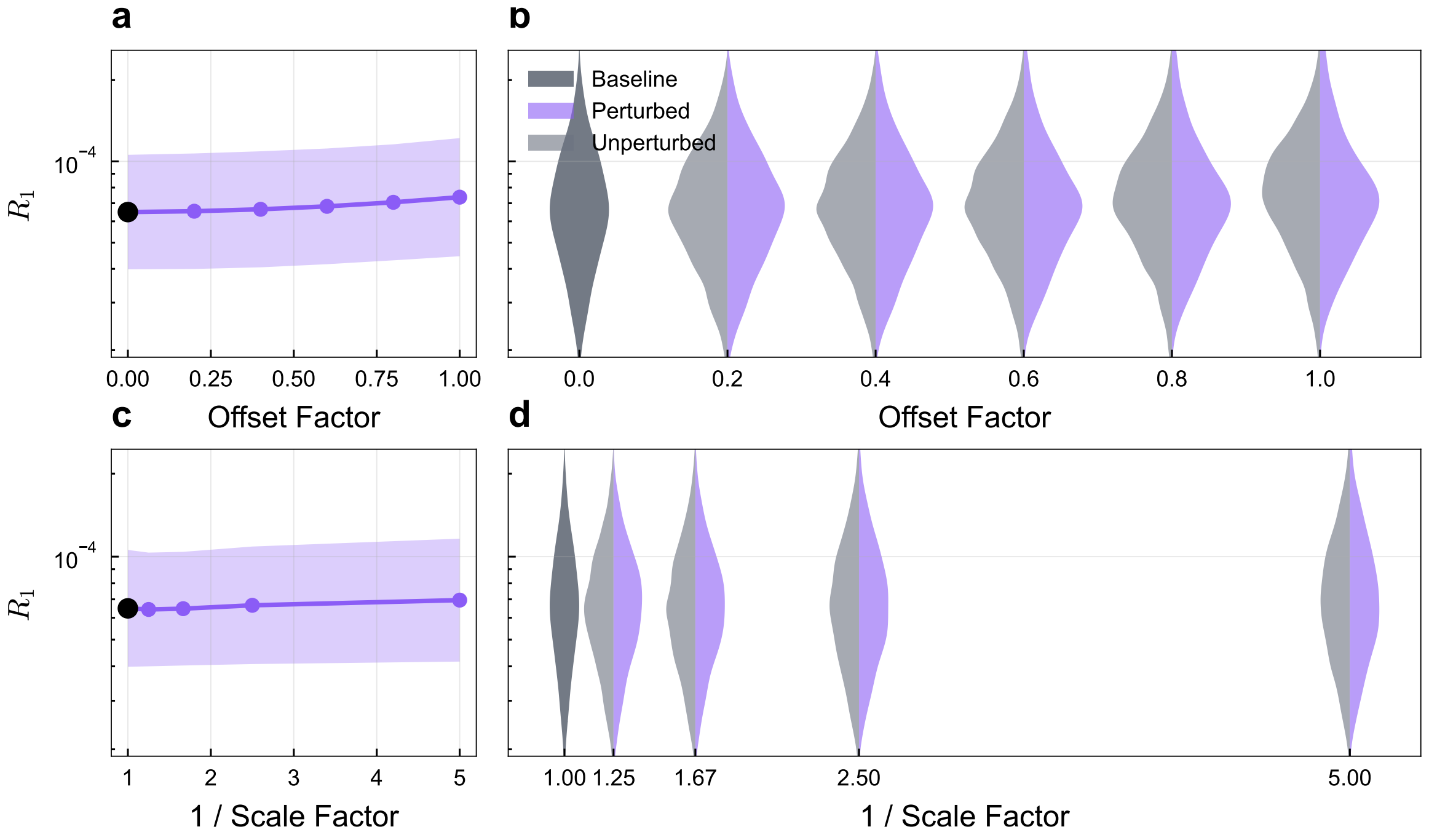


FIG. S12. Sorption residual $\mathscr{R}_1$ under input perturbations. (a, c) Mean $\mathscr{R}_1$ as a function of offset (a) and gain (c) perturbation magnitude. Shaded regions indicate the standard deviation. (b, d) Distribution of $\mathscr{R}_1$ at selected perturbation magnitudes for offset (b) and gain (d) conditions. Purple: perturbed channel; gray: unperturbed channels; dark gray: baseline (no perturbation).

edges. Since $\mathscr{R}_1$ carries one component per gas species, the per-channel detector takes the maximum over the two components: a junction is flagged if the sorption dynamics of either species are violated. For substitution splices, the median junction peak exceeds this baseline in all twelve channels for both $\mathscr{R}_1$ and $\mathscr{R}_2$ (weakest channel ratio 1.8), and taking the best channel per record detects the edit in 50 of 50 test sequences with a median peak-to-baseline ratio of 8.6. Deletion splices produce a milder dynamical break, since both sides of the junction originate from the same sequence. Per-channel margins therefore vary with the receptor coating (weakest median ratio 0.94), yet aggregating across channels still detects 48 of 50 sequences with a median best-channel ratio of 3.4. Junction detection is thus not specific to the channel shown in the main text, and it benefits directly from the channel diversity of the array.

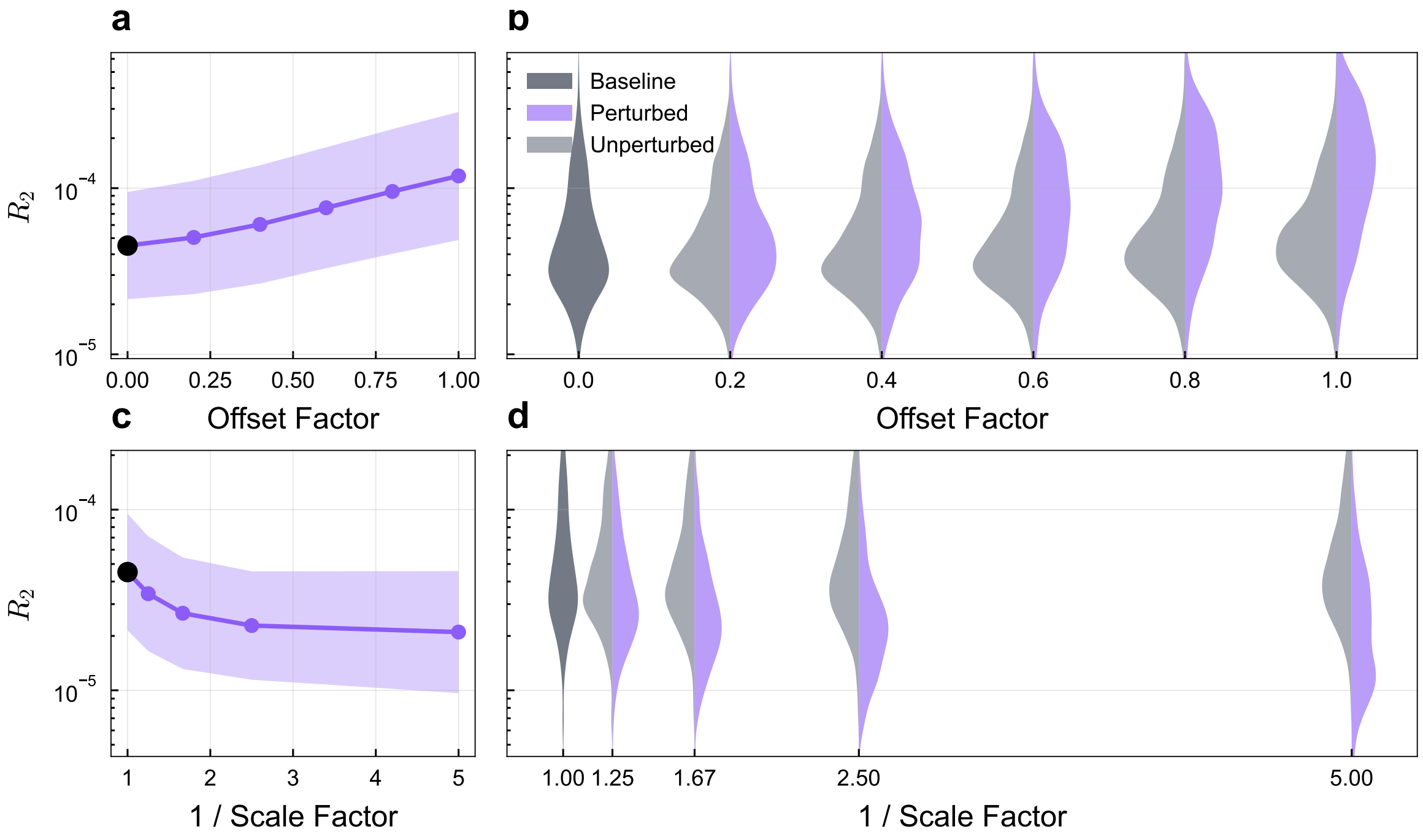


FIG. S13. Viscoelastic residual $\mathscr{R}_2$ under input perturbations, same layout as Fig. S12. $\mathscr{R}_2$ remains comparatively stable across both perturbation types, in contrast to the monotonic increase of $I_\sigma$.

## Supplementary Note 5. QUANTITATIVE RECONSTRUCTION METRICS

The full Physics model is evaluated on the complete test set (20 synchronous + 30 asynchronous sequences). Figure S17 shows the root-mean-square error (RMSE) for ethanol and water concentration reconstruction. Figure S18 presents the per-channel cumulative RMSE of signal reconstruction, and Fig. S19 shows the $R^2$ distribution across channels.

## Supplementary Note 6. BOUNDARY CASES AND FAILURE MODES

This section examines model behavior under extreme out-of-distribution (OOD) inputs that fall well outside the training distribution.

### 6.1. Sign-flip input

A sign-flipped input (all channels multiplied by $-1$ in normalized space) represents an extreme distributional shift (Fig. S20). The model produces erratic $\hat{C}_g$ predictions and large $I_\sigma$, as expected.

The $\Delta C_f$ trajectories show irregular patterns inconsistent with physical sorption dynamics. This extreme OOD condition serves as a sanity check: the model does not produce confident predictions on grossly invalid inputs, and the self-verification metrics correctly indicate unreliability.

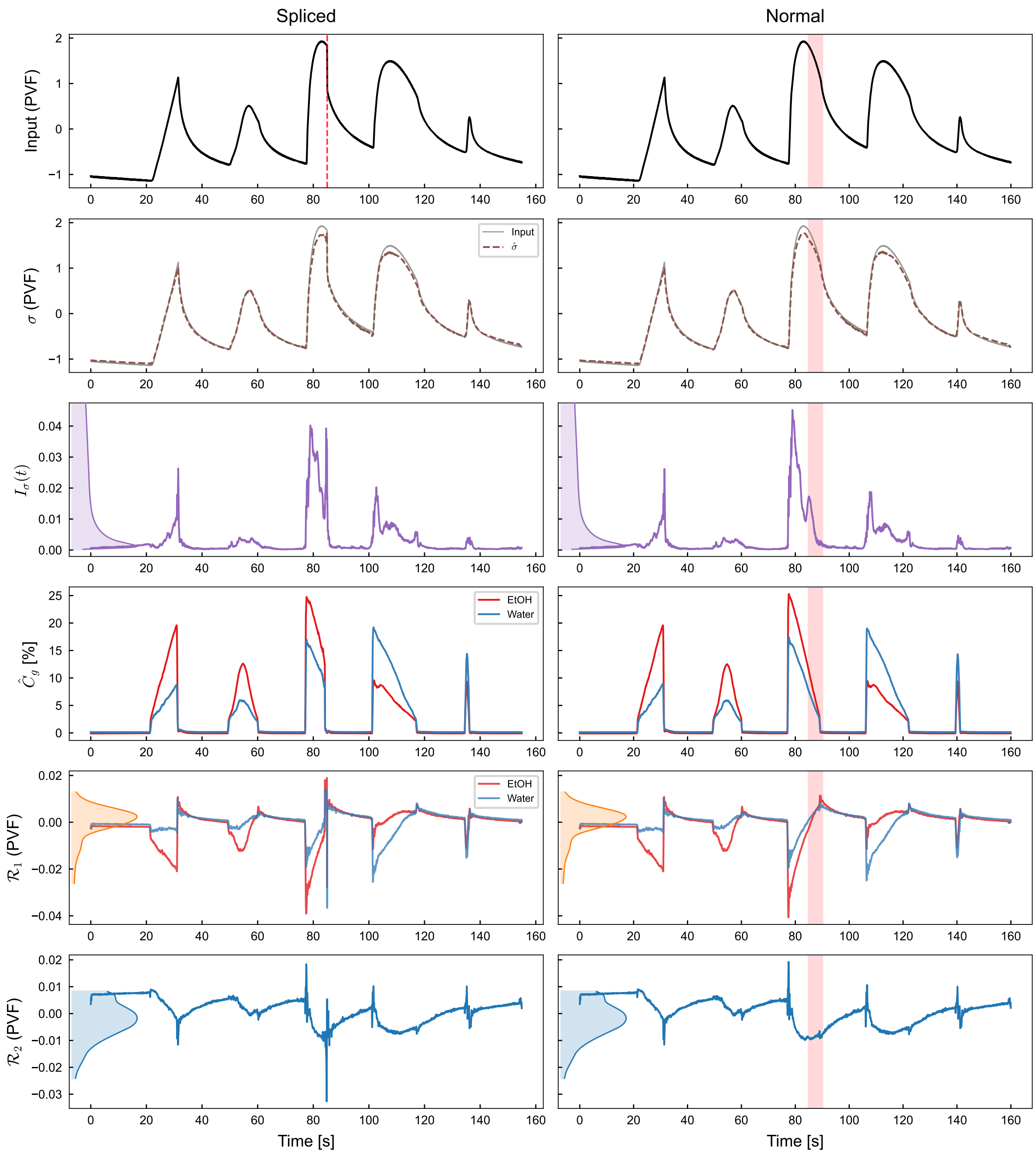


FIG. S14. Same-sequence deletion splice demonstrating the complementarity of $I_\sigma$ and physical residuals. Left column: spliced sequence (100 points removed at $t = 85$ s, red dashed line). Right column: normal reference (red shaded band indicates the removed segment). Rows from top: input signal, reconstructed $\hat{\sigma}$, time-resolved $I_\sigma(t)$, reconstructed $\hat{C}_g$, sorption residual $\mathscr{R}_1$, and viscoelastic residual $\mathscr{R}_2$. Inset KDE distributions at $t = 0$ show the baseline distributions from normal test data. The splice point produces sharp spikes in $\mathscr{R}_1$ and $\mathscr{R}_2$ but no appreciable change in $I_\sigma$.

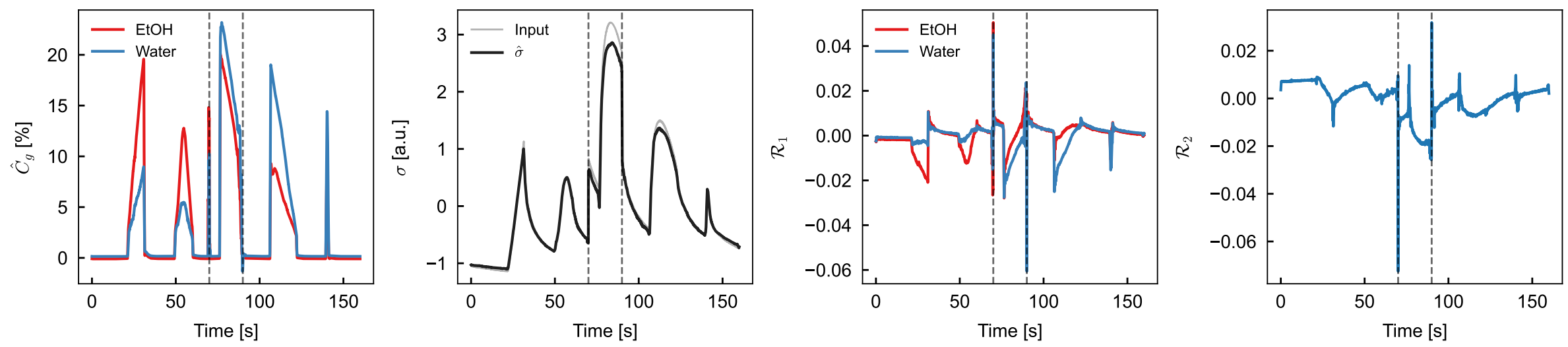


FIG. S15. Representative model outputs for a cross-sequence substitution splice. The 70–90 s segment is replaced by the corresponding segment of another test sequence; dashed lines mark the two junctions. From left: inferred gas-phase concentrations $\hat{C}_g$, substituted input versus reconstructed $\hat{\sigma}$ for a representative channel, sorption residual $\mathscr{R}_1$, and viscoelastic residual $\mathscr{R}_2$. The residuals spike at both junctions while $\hat{\sigma}$ shows no visible mismatch.

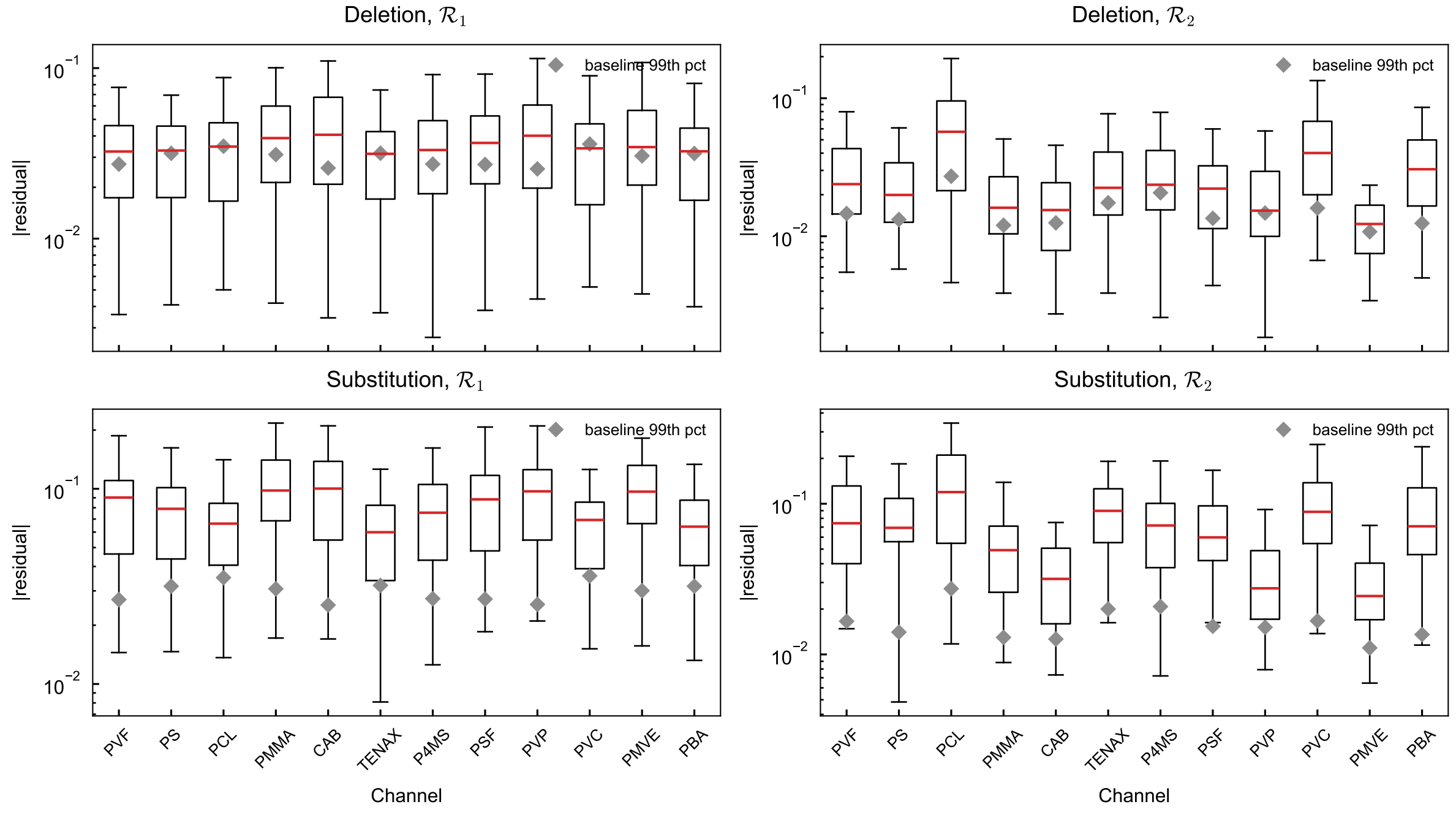


FIG. S16. Per-channel junction-detection statistics for deletion (top row) and substitution (bottom row) splices. Boxes show the distribution over the 50 test sequences of the peak absolute residual within ±1 s of the junction(s), for the sorption residual $\mathscr{R}_1$ (maximum over the two gases, left column) and the viscoelastic residual $\mathscr{R}_2$ (right column). Gray diamonds mark the median over sequences of the 99th percentile of the absolute residual over the remainder of each record. Note the logarithmic scale.

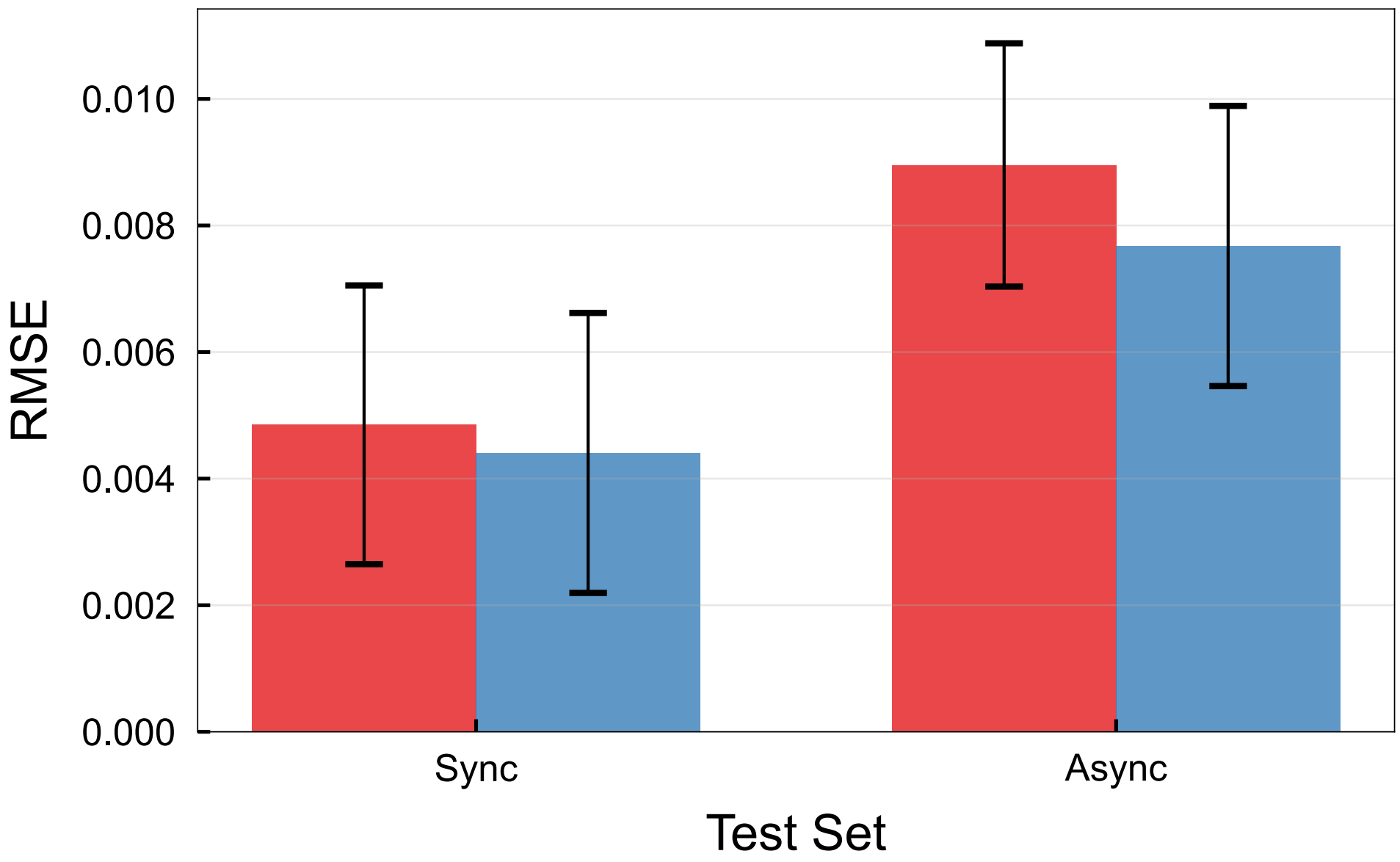


FIG. S17. RMSE of ethanol and water concentration reconstruction across the test set. Asynchronous mixing (left) and synchronous mixing (right). Error bars indicate the standard deviation.

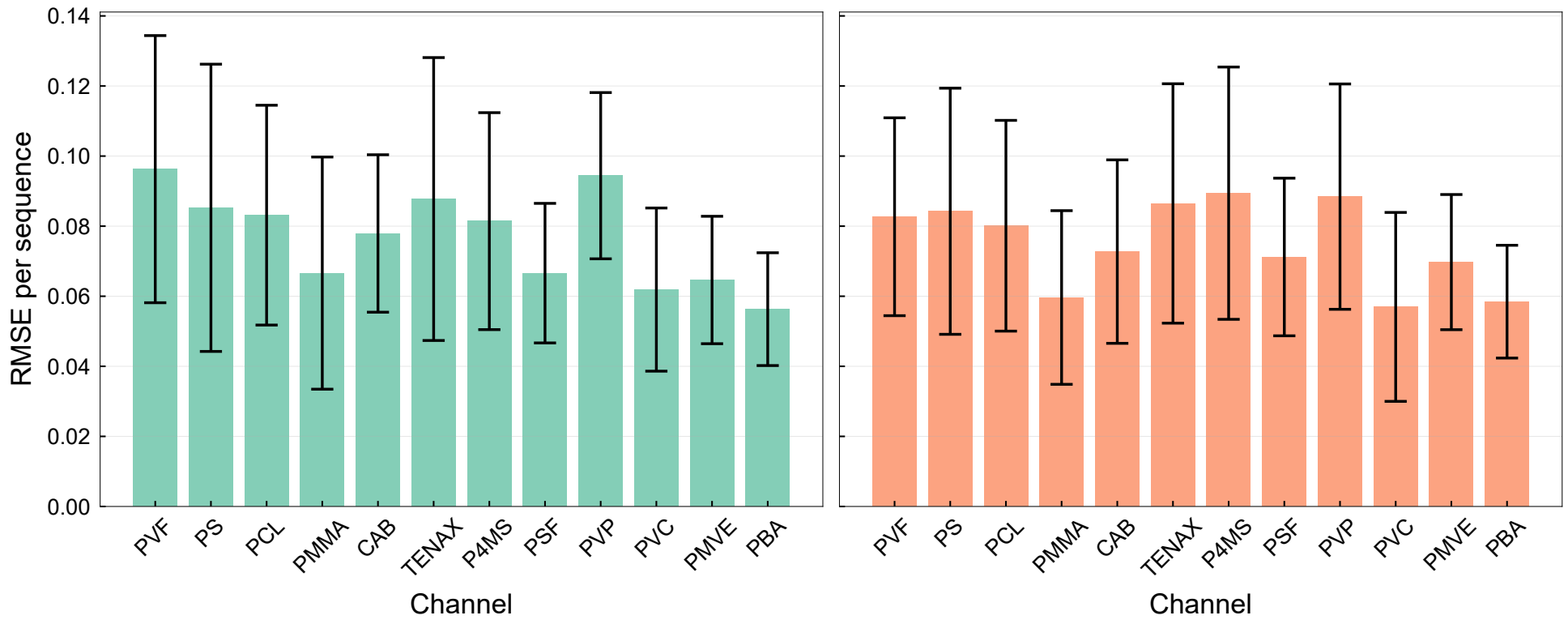


FIG. S18. Cumulative RMSE of reconstructed signal for each of the 12 sensor channels. Asynchronous mixing (left) and synchronous mixing (right). Error bars indicate the standard deviation.

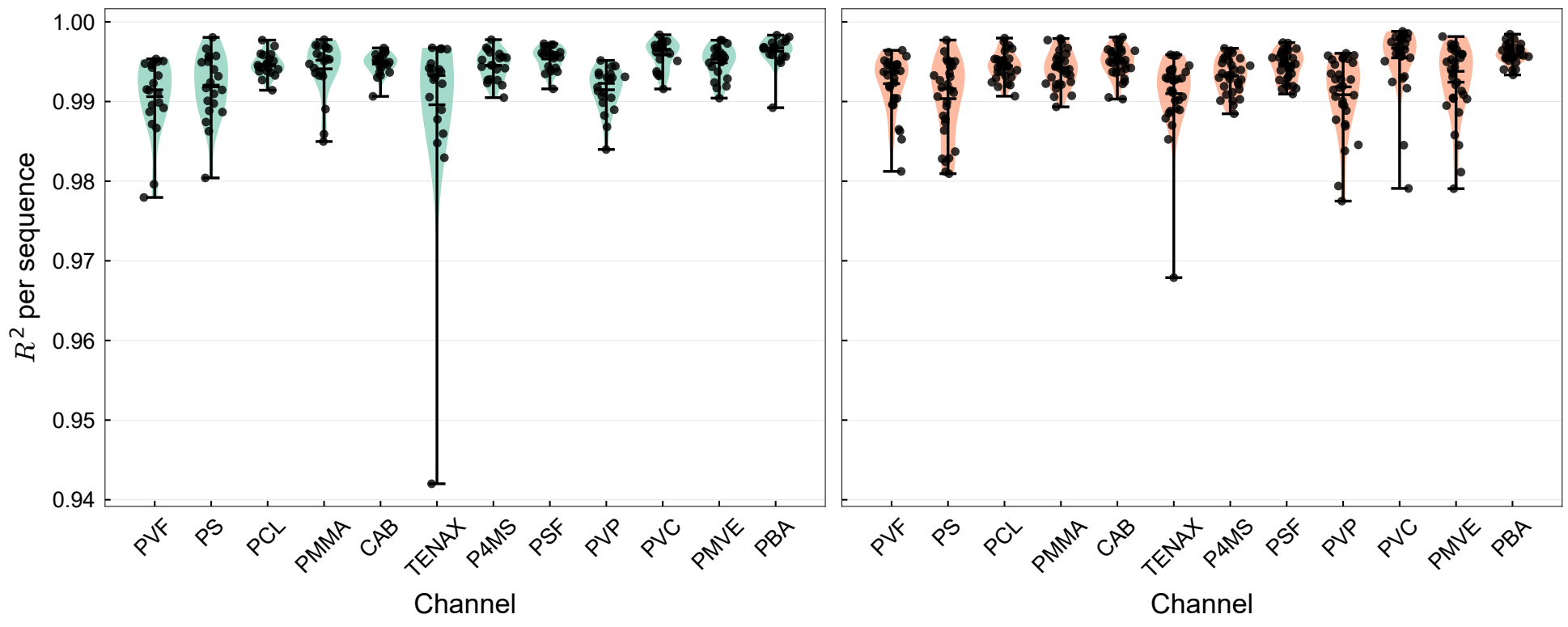


FIG. S19. Distribution of $R^2$ for the reconstructed signal across the 12 sensor channels. Asynchronous mixing (left) and synchronous mixing (right). Violin plots show the full distribution; markers indicate the extreme values.

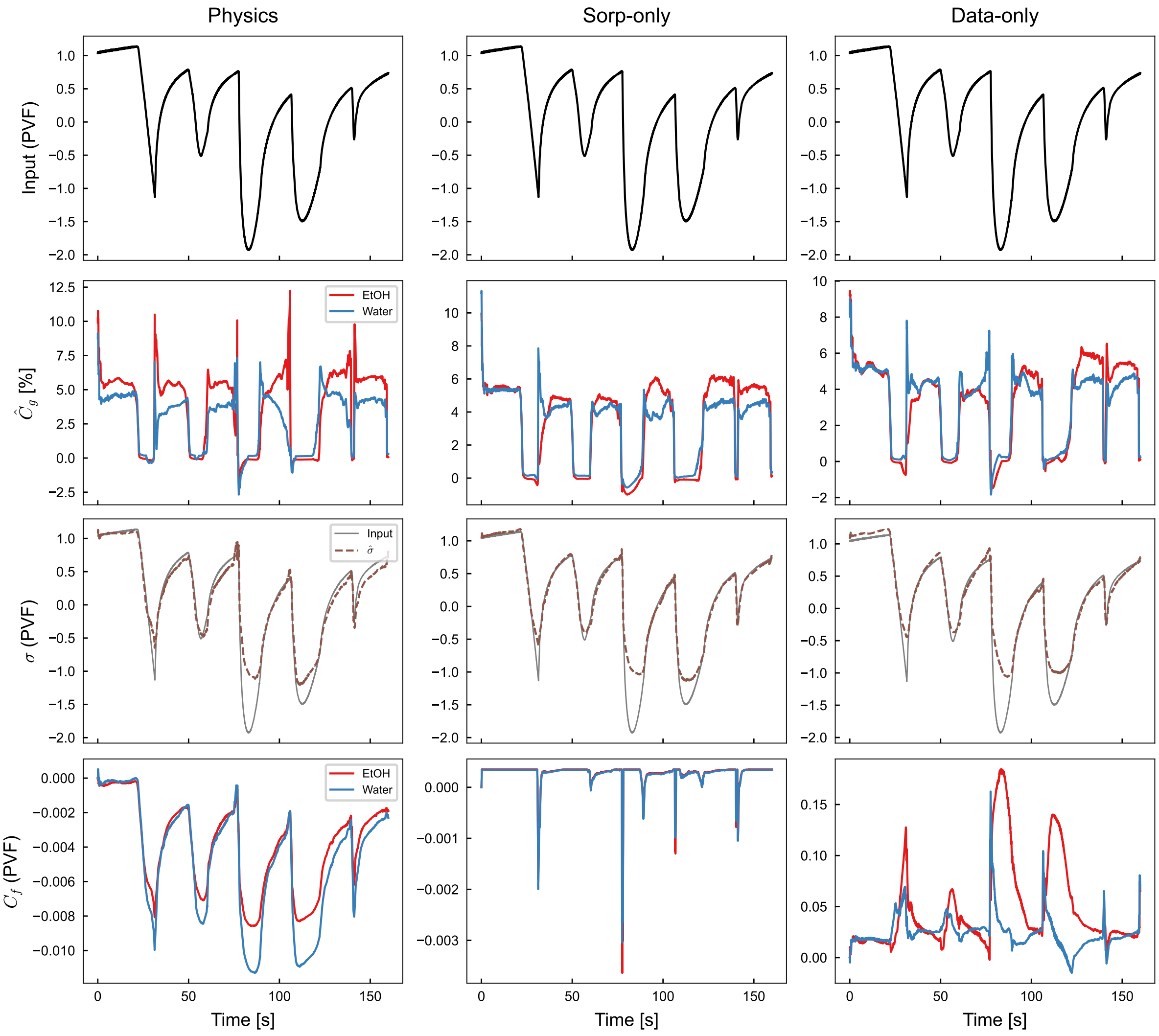


FIG. S20. Sign-flipped input (all channels multiplied by $-1$). Rows from top: input signal, reconstructed $\hat{C}_g$, reconstructed $\hat{\sigma}$ versus input, and $\Delta C_f$ for one representative channel (PVF).

## REFERENCES


[1]S. Bai, J. Z. Kolter, and V. Koltun, "An Empirical Evaluation of Generic Convolutional and Recurrent Networks for Sequence Modeling," (2018), arXiv:1803.01271 [cs].

[2]M. Almanstötter, R. Vetter, and D. Iber, "PINNverse: Accurate parameter estimation in differential equations from noisy data with constrained physics-informed neural networks," (2025), arXiv:2504.05248 [cs].

[3]X. Xie, P. Zhou, H. Li, Z. Lin, and S. Yan, "Adan: Adaptive Nesterov Momentum Algorithm for Faster Optimizing Deep Models," IEEE Trans. Pattern Anal. Mach. Intell. **46**, 9508–9520 (2024).

[4]G. Imamura, K. Shiba, G. Yoshikawa, and T. Washio, "Analysis of nanomechanical sensing signals; physical parameter estimation for gas identification," AIP Adv. **8**, 075007 (2018).

[5]K. Minami, K. Shiba, and G. Yoshikawa, "Sorption-induced static mode nanomechanical sensing with viscoelastic receptor layers for multistep injection-purge cycles," J. Appl. Phys. **129**, 124503 (2021).

[6]M.-Q. Feng, K. Minami, Y. Zhou, and G. Yoshikawa, "Analytical modeling and decoupling of humidity effects in nanomechanical sensing based on sorption kinetics and viscoelastic stress relaxation," Phys. Rev. E **111**, 065407 (2025).

[7]K. Minami and G. Yoshikawa, "An Analytical Model of Sorption-Induced Static Mode Nanomechanical Sensing for Multicomponent Analytes," Anal. Chem. **97**, 19306–19312 (2025).

[8]S. Xu, C. Yan, Z. Sun, R. Huang, D. Guo, and G. Yang, "On the Preprocessing of Physics-informed Neural Networks: How to Better Utilize Data in Fluid Mechanics," (2024), arXiv:2403.19923 [physics].



\end{document}